\documentclass[amsmath,amssymb,aps,prfluids]{revtex4-2}
\usepackage{dcolumn}
\usepackage{bm}
\usepackage{graphicx} 
\usepackage{epstopdf}
\usepackage[titletoc,title]{appendix}
\usepackage{graphicx}
\usepackage{dcolumn}
\usepackage{bm}
\usepackage{xcolor}
\usepackage{comment}
\usepackage[normalem]{ulem}

\begin{document}

\preprint{APS/123-QED}

\title{Slow linear shear flow past discrete particles adhered to a plane wall}

\author{Itzhak Fouxon$^{1,2}$}   \author{Alexander M. Leshansky$^1$}\email{lisha@technion.ac.il}
\affiliation{$^1$Department of Chemical Engineering, Technion, Haifa 32000, Israel}
\affiliation{$^2$Department of Computational Science and Engineering, Yonsei University, Seoul 120-749, South Korea}

\begin{abstract}

Linear shear flow bounded by a plane wall is an idealization that occurs in microfluidic devices and many other applications. Perfect plane approximation neglects surface irregularities and discrete particles adsorbed at the surface. Here we study the disturbance to the linear shear flow due to the particle(s) rigidly attached to the surface. We first revisit the exact solution of O'Neill \cite{o1968sphere} for a spherical particle in contact with an infinite plane boundary. While the original paper contains multiple typos and provides very few details of the derivation, we present detailed solution accompanied by an alternative and simpler derivation of the viscous force and the torque exerted on the particle. We further study the universal far-field behavior of the flow due to an \emph{arbitrary} particle adhered to the surface, and demonstrate that it is controlled by the stress moment of magnitude depending on particle's volume and shape. Using the revised O'Neill solution, we compute the stress moment for a spherical particle. Using the far-field asymptotic form of the flow we estimate the net flow due to uniform and sparse layer of discrete adsorbates by superposition and demonstrate that it does not decay away from the plane. 

\end{abstract}

\maketitle

\section{Introduction \label{sec:introduction}}

Wall-bounded shear flow is one of the most basic flows of low-Reynolds-number hydrodynamics. One of the most common examples is the pressure-driven (Poiseuille) flow near the wall of the microfluidic slit channel.  
The flow is parallel to the plane wall and its magnitude grows linearly with the distance from the no-slip surface on which the flow velocity vanishes. However, realistic surfaces are typically non-smooth due to intrinsic surface roughness and/or presence of discrete micro- and nanometric particles adsorbed onto the surface from the liquid. These irregularities may distort the idealized unidirectional shear flow, possibly generating a rather complex flow pattern. The intuitive assumption would be that the flow bounded by a realistic non-smooth surface can be approximated by that above an ideally smooth plane wall at distances much further than the characteristic linear scale of the surface corrugation or size of the discrete adsorbates. In other words, one might anticipate the flow \emph{disturbance} to be confined to the near vicinity of the bounding surface depth comparable to a typical length scale of the roughness or diameter of the adsorbates. However low-Reynolds-number flows are governed by the Stokes equations, that are rather similar to the Laplace equation of electrostatics. Considering the non-decaying electric field induced by the surface charge distributed at the flat boundary, one might anticipate the emergence of the (non-decaying) secondary flow away from the bounding wall due to surface corrugations or a layer of discrete absorbates. The well-known example of such secondary flow is the microfluidic mixer based on patterns of shallow grooves fabricated on the floor of the microchannel using photolithography \cite{stroock2002chaotic}.

In the present paper we focus on flow disturbance due to discrete adsorbates and consider the analytically trackable case of a linear shear flow past to a spherical particle in a contact with a rigid plane. The exact solution in special``touching sphere" coordinates was first reported by O'Neill \cite{o1968sphere}, however the paper contains multiple typos that are difficult to detect, as very little details of the derivation are provided. We, therefore, revisit the solution and provide the detailed derivations correcting the typos in Ref.~\cite{o1968sphere}. The expressions for the viscous force and torque exerted on the particle by the flow are obtained using an alternative and simpler approach that involves integration over the plane and then confirmed by the direct integration over the particle surface.  

We then extend the analysis to adsorbates of an \emph{arbitrary} shape, using integral representation of the wall-bounded flow. This allows to derive a universal closed-form expression for the far-field flow disturbance, being controlled by the scalar stress moment $M$ of magnitude depending on the particle volume and shape. The value of $M$ is then computed explicitly for a spherical particle using the revised O'Neill solution. Based on the far-field asymptotic form of the flow disturbance due to a single particle, we then estimate the \emph{constant} net flow induced by a uniform layer of adsorbates with low surface coverage using superposition. 

The present study is also relevant to the analysis of the signal (impedance) of liquid-phase Quart Crystal Microbalance (QCM) device, that relies on the fact that small matter adsorbed on the surface of the fast oscillating crystal, changes the frequency and the decay rate of the oscillations \cite{review2015}. Although the frequency of the oscillations is in MHz range, for nanometric adsorbates (e.g., such as viruses or protein molecules that are much smaller than the viscous penetration depth, $\delta\approx 250$~nm in water at the fundamental frequency of 5~MHz), the corresponding hydrodynamic problem in the frequency domain reduces to that studied here \cite{leshansky2024quartz}.

\section{Shear flow past a spherical particle attached to a plane revisited}

\subsection{Problem formulation} \label{s}

\begin{figure}
    \includegraphics[width=0.4\textwidth]{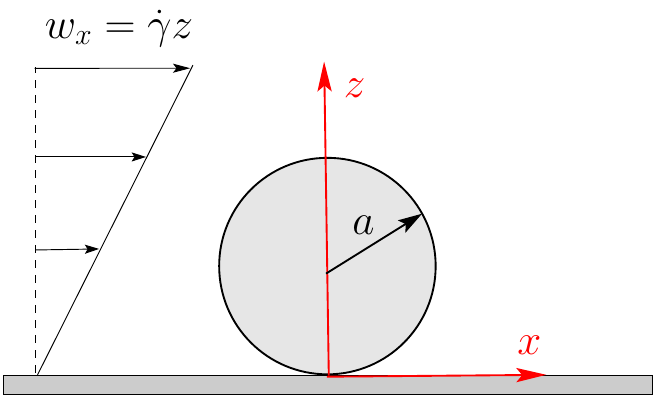}
    \caption{\label{fig:schematic} Schematic illustration of the problem setup. A fixed rigid spherical particle of
radius $a$ is in contact with a rigid plane at $z\!=\!0$ in which the origin is at the point of contact. The viscous fluid above the plane
is set into motion with the velocity $w_x=\dot{\gamma} x$ (without the particle). 
    }
\end{figure}

We consider the wall-bounded shear flow of the incompressible Newtonian fluid, $\bm w=(\dot{\gamma} z, 0, 0)$, where $\dot{\gamma}$ is the shear-rate,  impinging on the spherical particle with radius $a$ rigidly attached to an ideally smooth plane at the point $z\!=\!0$ (see Fig.~\ref{fig:schematic}). 
We use dimensionless variables, where all lengths are scaled with the particle radius $a$, the fluid velocity with $a\dot{\gamma}$ and the pressure $p$ with $\eta\dot{\gamma}$, where $\eta$ is the dynamic viscosity. The flow is governed by the Stokes and continuity equations accompanied by the no-slip boundary conditions on all surfaces: 
\begin{eqnarray}&&\!\!\!\!\!\!\!\!\!
\nabla p=\nabla^2\bm w, \ \ \nabla\cdot \bm w=0,\quad \mbox{and} \quad \bm w(z=0)=0,\ \ \bm w(|\bm q|\to\infty)=z\hat{\bm x},\ \ \bm w(|\bm q|=1)=\bm 0.
\end{eqnarray}
The center of the sphere has coordinates $(0, 0, 1)$ and $\bm q$ is the corresponding radius-vector with the origin at the particle center, $\bm q\equiv (x, y, z-1)$. The coordinate origin is placed at the point of contact.

The solution can be written as $\bm w=z\hat{\bm x}
+\bm v$ where the flow perturbation $\bm v$ obeys 
\begin{eqnarray}&&\!\!\!\!\!\!\!\!\!
\nabla p=\nabla^2\bm v, \ \ \nabla\cdot \bm v=0,\ \quad \mbox{and} \quad \bm v(z=0)=0,\  \bm v(|\bm q|\to\infty)=\bm 0,\ \ 
\bm v(|\bm q|=1)=-z \hat{\bm x}.  \label{eqv}
\end{eqnarray}
In the cylindrical coordinates $\{\rho, \varphi, z\}$ the Eqs.~(\ref{eqv}) take the form
\begin{eqnarray}&&\!\!\!\!\!\!\!\!\!
\frac{\partial p}{\partial \rho}=\nabla^2 v_{\rho}-\frac{v_{\rho}}{\rho^2}-\frac{2}{\rho^2}\frac{\partial v_{\varphi}}{\partial \varphi}, \ \quad 
\frac{1}{\rho}\frac{\partial p}{\partial \varphi}=\nabla^2 v_{\varphi}-\frac{v_{\varphi}}{\rho^2}+\frac{2}{\rho^2}\frac{\partial v_\rho}{\partial \varphi}, \ \quad  \frac{\partial p}{\partial z}=\nabla^2 v_z,\nonumber 
\\&& \!\!\!\!\!\!\!\!\!\frac{1}{\rho}\frac{\partial (\rho v_{\rho})}{\partial \rho}+\frac{1}{\rho}\frac{\partial v_{\varphi}}{\partial \varphi}+\frac{\partial v_z}{\partial z}=0,\ 
\nonumber
\end{eqnarray}
with the boundary conditions:
\begin{eqnarray}&&
v_{\rho}=-z\cos\varphi\,,\ \ v_{\varphi}=z\sin\varphi\,, \ v_z=0 \quad \mbox{at} \quad |\bm q|=1 , \ \ 
\bm v(z=0)=0,\ \ \bm v(|\bm q|\to\infty)={\bm 0}. \nonumber
\end{eqnarray}
The dependence on $\varphi$ is resolved by the ansatz
\begin{eqnarray}&&\!\!\!\!\!\!\!\!\!
v_{\rho}=U(\rho, z)\cos\varphi, \ \ v_{\varphi}=V(\rho, z)\sin\varphi,  \ \ v_z=W(\rho, z)\cos\varphi,\ \ p=P(\rho, z)\cos\varphi,\label{ca}
\end{eqnarray}
where we introduced the scalar functions $U$, $V$, $W$, $P$ that depend on $\rho$ and $z$ only. To derive the equations obeyed by these functions, we follow \cite{o1968sphere} and express the solution of Eqs.~(\ref{eqv}) via four \textit{harmonic} functions given by the pressure and the components of the auxiliary \emph{harmonic} vector field $\bm u\equiv \bm v-p\bm r/2$,  
\begin{eqnarray}&&\!\!\!\!\!\!\!\!\!
u_{\rho}=\left(U(\rho, z)-\frac{\rho P(\rho, z)}{2}\right)\cos\varphi, \ \ u_{\varphi}=V(\rho, z)\sin\varphi,  \ \ u_z=\phi(\rho, z)\cos\varphi,\ \ \phi\equiv W(\rho, z)-\frac{z P(\rho, z)}{2}.\nonumber
\end{eqnarray}
It can be readily shown that 
\begin{eqnarray}&&\!\!\!\!\!\!\!\!\!
\nabla^2 p=0, \ \ \nabla^2\bm u=0. \label{Lapls}
\end{eqnarray}
Then by using $p=P\cos{\varphi}$ in $\nabla^2 p=0$ and $u_z=\phi\cos{\varphi}$ in $\nabla^2 u_z=0$ it follows that
\begin{eqnarray}&&\!\!\!\!\!\!\!\!\!
{\hat L}_1 P={\hat L}_1\phi=0,\ \ {\hat L}_m\equiv \nabla^2-\frac{m^2}{\rho^2}=\frac{\partial^2}{\partial \rho^2}+\frac{1}{\rho}\frac{\partial}{\partial \rho}-\frac{m^2}{\rho^2}+\frac{\partial^2}{\partial z^2},
\label{pressure}\end{eqnarray}
where the operators ${\hat L}_m$ are defined as in \cite{o1968sphere}. We further find that 
\begin{eqnarray}&&\!\!\!\!\!\!\!\!\! u_x=\phi(\rho, z)\cos^2\varphi-V(\rho, z)\sin^2\varphi,\ \quad \nabla^2 \langle u_x\rangle={\hat L}_0 \langle u_x\rangle=0, \label{angl}
\end{eqnarray}
where the angular brackets stand for averages over $\varphi$. This implies that 
\begin{eqnarray}&&\!\!\!\!\!\!\!\!\!
{\hat L}_0 \psi(\rho, z)=0;\ \quad 
\psi(\rho, z) \equiv U(\rho, z)-\frac{\rho P(\rho, z)}{2}-V(\rho, z).\label{lk}
\end{eqnarray}
Similarly, using $\nabla^2u_y=0$ we find the last of the four equations for the functions $U$, $V$, $W$ and $P$:
\begin{eqnarray}&&\!\!\!\!\!\!\!\!\! 
u_y=\chi(\rho, z)\sin\varphi\cos\varphi,\ \ {\hat L}_2\chi(\rho, z)=0;\ \ \chi(\rho, z)\equiv U(\rho, z)-\frac{\rho P(\rho, z)}{2}+V(\rho, z).
\end{eqnarray}
These equations could also be obtained directly by using the ansatz (\ref{ca}) in Eqs.~(\ref{eqv}), which gives
\begin{eqnarray}&&\!\!\!\!\!\!\!\!\!
\frac{\partial P}{\partial \rho}=\nabla^2 U-\frac{2(U+V)}{\rho^2}, \ \quad 
-\frac{P}{\rho}=\nabla^2 V-\frac{2(U+V)}{\rho^2}, \ \ \frac{\partial P}{\partial z}=\nabla^2 W-\frac{W}{\rho^2},\nonumber\\&& \!\!\!\!\!\!\!\!\!
\frac{1}{\rho}\frac{\partial (\rho U)}{\partial \rho}+\frac{V}{\rho}+\frac{\partial W}{\partial z}=0, \label{cont} 
\end{eqnarray} 
where we used $\nabla^2 (U\cos\varphi)=\cos\varphi \left(\nabla^2 U-U/\rho^2\right)$ and $\nabla^2 (V\sin\varphi)=\sin\varphi \left(\nabla^2 V-V/\rho^2\right)$ and the boundary conditions
\begin{eqnarray}
&& U=-z,\ V=z, \ W=0 \ \quad \mbox{at}  \quad|\bm q|=1,\ \ U=V=W=0 \quad  \mbox{at}  \quad z=0 \,. \label{a} 
\end{eqnarray}

Thus the problem reduces to solving for the zero modes of ${\hat L}_m$. 

\subsection{Zero modes of the operators ${\hat L}_m$}

Below we derive zero modes of the operator ${\hat L}_m$ in ``tangent sphere" coordinates defined by: 
\begin{eqnarray}&&\!\!\!\!\!\!\!\!\!
\eta(\rho, z)=\frac{2\rho}{\rho^2+z^2},\ \ \xi(\rho, z)=\frac{2z}{\rho^2+z^2};\ \quad \  \rho(\eta, \xi)=\frac{2\eta}{\xi^2+\eta^2},\ \ z(\eta, \xi)=\frac{2\xi}{\xi^2+\eta^2}. 
\end{eqnarray}
In this coordinate system the particle surface defined by $|\bm q|=\sqrt{\rho^2+(z-1)^2}=1$ is given by the isoline $\xi=1$, and the plane surface by $\xi=0$; the flow domain is $0\leq \xi\leq 1$ and $0\leq \eta< \infty$. The form of the operators ${\hat L}_m$ in these coordinates can be found from
\begin{eqnarray}&&\!\!\!\!\!\!\!\!\!
\frac{\partial}{\partial \rho}
=\frac{\xi^2-\eta^2}{2}\frac{\partial}{\partial \eta}-\eta \xi\frac{\partial}{\partial \xi};\ \ 
\frac{\partial}{\partial z}
=\frac{\eta^2-\xi^2}{2}\frac{\partial}{\partial \xi}-\eta \xi\frac{\partial}{\partial \eta},
\end{eqnarray}
implying ($\partial_z^2$ can be obtained from $\partial_{\rho}^2$ by interchanging $\xi$ and $\eta$)
\begin{eqnarray}&&\!\!\!\!\!\!\!\!\!
\frac{\partial^2}{\partial \rho^2}=\frac{(\xi^2-\eta^2)^2}{4}\frac{\partial^2}{\partial \eta^2}-\frac{\eta(\xi^2-\eta^2)}{2}\frac{\partial}{\partial \eta}+\eta^2 \xi^2\frac{\partial^2}{\partial \xi^2}+\eta^2 \xi\frac{\partial}{\partial \xi}
-\eta \xi(\xi^2-\eta^2)\frac{\partial^2}{\partial \eta\partial \xi}
-\frac{\xi(\xi^2-\eta^2)}{2}\frac{\partial}{\partial \xi}-\eta \xi^2\frac{\partial}{\partial \eta},\nonumber\\&&\!\!\!\!\!\!\!\!\!
\frac{\partial^2}{\partial z^2}=\frac{(\eta^2-\xi^2)^2}{4}\frac{\partial^2}{\partial \xi^2}-\frac{\xi(\eta^2-\xi^2)}{2}\frac{\partial}{\partial \xi}+\xi^2 \eta^2\frac{\partial^2}{\partial \eta^2}+\xi^2 \eta\frac{\partial}{\partial \eta}
-\eta \xi(\eta^2-\xi^2)\frac{\partial^2}{\partial \eta\partial \xi}
-\frac{\eta(\eta^2-\xi^2)}{2}\frac{\partial}{\partial \eta}-\eta^2 \xi\frac{\partial}{\partial \xi}.
\end{eqnarray}
We next find that 
\begin{eqnarray}&&\!\!\!\!\!\!\!\!\!
\frac{\partial^2}{\partial \rho^2}+\frac{\partial^2}{\partial z^2}
=\frac{(\xi^2+\eta^2)^2}{4}\left(\frac{\partial^2}{\partial \eta^2}+\frac{\partial^2}{\partial \xi^2}\right),\nonumber\\
&&\!\!\!\!\!\!\!\!\! {\hat L}_m=\frac{(\xi^2+\eta^2)^2}{4}\left(\frac{\partial^2}{\partial \eta^2}+\frac{\partial^2}{\partial \xi^2}\right)
+\frac{\xi^2+\eta^2}{2\eta}\left(\frac{\xi^2-\eta^2}{2}\frac{\partial}{\partial \eta}-\eta \xi\frac{\partial}{\partial \xi}\right)
-\frac{m^2(\xi^2+\eta^2)^2}{4\eta^2}.
\end{eqnarray}
The following operator identities 
\begin{eqnarray}&&\!\!\!\!\!\!\!\!\!
\left(\frac{\partial^2}{\partial \eta^2}+\frac{\partial^2}{\partial \xi^2}\right)\sqrt{\xi^2+\eta^2}
=\sqrt{\xi^2+\eta^2}\left(\frac{\partial^2}{\partial \eta^2}+\frac{\partial^2}{\partial \xi^2}\right)
+\frac{2\xi}{\sqrt{\xi^2+\eta^2}}\frac{\partial}{\partial \xi}+\frac{2\eta}{\sqrt{\xi^2+\eta^2}}\frac{\partial}{\partial \eta}+\frac{1}{\sqrt{\xi^2+\eta^2}}
\nonumber\\&&\!\!\!\!\!\!\!\!\!
\left(\frac{\xi^2-\eta^2}{2}\frac{\partial}{\partial \eta}-\eta \xi\frac{\partial}{\partial \xi}\right)\sqrt{\xi^2+\eta^2}=\sqrt{\xi^2+\eta^2}\left(\frac{\xi^2-\eta^2}{2}\frac{\partial}{\partial \eta}-\eta \xi\frac{\partial}{\partial \xi}\right)
-\frac{\eta\sqrt{\xi^2+\eta^2}}{2},\nonumber
\end{eqnarray}
imply the general relation
\begin{eqnarray}&&\!\!\!\!\!\!\!\!\!
{\hat L}_m \sqrt{\xi^2+\eta^2} 
=\frac{(\xi^2+\eta^2)^{5/2}}{4}\left(\frac{\partial^2}{\partial \eta^2}+\frac{\partial^2}{\partial \xi^2}
+\frac{1}{\eta}\frac{\partial}{\partial \eta}
-\frac{m^2}{\eta^2}\right).
\end{eqnarray}
We find from Eq.~(\ref{pressure}) that the pressure field satisfies 
\begin{eqnarray}&&\!\!\!\!\!\!\!\!\!
\frac{\partial^2 f}{\partial \eta^2}
+\frac{1}{\eta}\frac{\partial f}{\partial \eta}
-\frac{f}{\eta^2}=-\frac{\partial^2 f}{\partial \xi^2};
\label{pr}
\end{eqnarray} 
where we introduced $f \equiv P/\sqrt{\xi^2+\eta^2}$. We next apply the Hankel transform of order one defined as  
\begin{eqnarray}&&\!\!\!\!\!\!\!\!\!
\widehat f(k, \xi)=k\int_0^{\infty} f(\eta, \xi)J_{1}(k\eta)\eta d\eta,\ \ f(\eta, \xi)=\int_0^{\infty} \widehat f(k, \xi)J_{1}(k\eta) dk.
\end{eqnarray} 
to the LHS of Eq.~(\ref{pr}) to obtain
\begin{eqnarray}&&\!\!\!\!\!\!\!\!\!
k\int_0^{\infty} d\eta J_{1}(k\eta)\eta\left(\frac{\partial^2 f}{\partial \eta^2}
+\frac{1}{\eta}\frac{\partial f}{\partial \eta}
-\frac{f}{\eta^2}\right)=k\int_0^{\infty} \eta f d\eta\left(\frac{\partial^2}{\partial \eta^2}
+\frac{1}{\eta}\frac{\partial }{\partial \eta}
-\frac{1}{\eta^2}\right) J_{1}(k\eta)=-k^2 \widehat{f}(k, \xi),
\end{eqnarray} 
where we used $J_1(0)=0$. Thus, 
$\widehat f(k, \xi)$ obeys 
\begin{eqnarray}&&\!\!\!\!\!\!\!\!\!
\frac{\partial^2 {\widehat f}}{\partial \xi^2}=k^2 {\widehat f},\  \quad \widehat f(k, \xi)=B(k)\cosh(k\xi)+C(k)\sinh(k\xi),
\end{eqnarray} 
where $B(k)$ and $C(k)$ are some arbtrary functions of $k$. We remark that Hankel transform of an integer order $\nu$ is equivalent to rescaled $2(\nu+1)-$dimensional Fourier transform of $f(\eta, \xi)$, considered as an angle-independent function in $2(\nu+1)$ dimensions with $\eta$ the radial variable. This implies that the above solution is valid provided that $f(\eta, \xi)$, as a function of $\eta$, obeys the well-known conditions for the Fourier transform. 

Similar considerations give
\begin{eqnarray}&&\!\!\!\!\!\!\!\!\!
\widehat g(k, \xi)={\widetilde A}(k)\cosh(k\xi)+A(k)\sinh(k\xi), \ \ g(\eta, \xi)\equiv\frac{\phi(\eta, \xi)}{\sqrt{\xi^2+\eta^2}},
\end{eqnarray} 
where $A(k)$, ${\widetilde A}(k)$ are some functions of $k$. Using the boundary condition $W(\xi=0)=0$ we have ${\widehat g}(k, \xi=0)=0$ which gives ${\widetilde A}(k)=0$.  Similar considerations hold for other velocity components, for which we have: 
\begin{eqnarray}&&\!\!\!\!\!\!\!\!\!
\left(\frac{\partial^2}{\partial \eta^2}\!+\!\frac{\partial^2}{\partial \xi^2}
\!+\!\frac{1}{\eta}\frac{\partial}{\partial \eta}\right)\frac{\psi(\eta, \xi)}{\sqrt{\eta^2\!+\!\xi^2}}\!=\!0,\ \quad \left(\frac{\partial^2}{\partial \eta^2}\!+\!\frac{\partial^2}{\partial \xi^2}
\!+\!\frac{1}{\eta}\frac{\partial}{\partial \eta}
\!-\!\frac{4}{\eta^2}\right)\frac{\chi(\eta, \xi)}{\sqrt{\eta^2\!+\!\xi^2}}\!=\!0,
\end{eqnarray}
whose solutions are found by using the Hankel transforms of the zeroth and second order, respectively, where 
\begin{eqnarray}&&\!\!\!\!\!\!\!\!\!
\widehat h(k,\xi)=k\int_0^{\infty} h(\eta,\xi) J_{\nu}(k\eta)\eta d\eta,\ \quad h(\eta,\xi)=\int_0^{\infty} \widehat h(k,\xi) J_{\nu}(kr) dk,
\end{eqnarray}
with $\nu$ the order of the transform. Finally, the solutions are given by \cite{o1968sphere}
\begin{eqnarray}&&\!\!\!\!\!\!\!
P\! \equiv 2Q=\!\sqrt{\xi^2\!+\!\eta^2}\int_0^{\infty} \!\!\left(B(k)\cosh(k\xi)\!+\!C(k)\sinh(k\xi)\right)J_{1}(k\eta) dk;\ \quad \phi=\!\sqrt{\xi^2\!+\!\eta^2}\int_0^{\infty}\!\! A(k)\sinh(k\xi)J_{1}(k\eta)dk, \label{oneil}  \\&&\!\!\!\!\!\!\! \psi\!=\!\sqrt{\xi^2\!+\!\eta^2}\!\!\int_0^{\infty} \!\!\!\left(D(k)\cosh(k\xi)\!+\!E(k)\sinh(k\xi)\right)\! J_{0}(k\eta) dk;\,  \chi\!=\!\sqrt{\xi^2\!+\!\eta^2}\!\!\int_0^{\infty} \!\!\!\left(F(k)\cosh(k\xi)\!+\!G(k)\sinh(k\xi)\right)\! J_{2}(k\eta) dk. \nonumber
\end{eqnarray}
The $7$ unknown functions of $k$ have to be determined from the continuity equations and the no-slip boundary conditions. 

\subsection{Incompressibility}

The continuity Eq.~(\ref{cont}) can be written as
\begin{eqnarray}&&\!\!\!\!\!\!\!\!\! 
0=2\left(\frac{\partial U}{\partial \rho}\!+\!\frac{U+V}{\rho}\!+\!\frac{\partial W}{\partial z}\right)
\!=\!2\left(3\!+\!\rho\frac{\partial}{\partial\rho}\!+\!z\frac{\partial}{\partial z}\right)Q
\!+\!\frac{\partial \chi}{\partial \rho}\!+\!\frac{\partial \psi}{\partial \rho}\!+\!\frac{2\chi}{\rho}\!+\!2\frac{\partial\phi}{\partial z}.\label{incomp}
\end{eqnarray}
It simplifies the calculations to notice that some terms in the RHS of (\ref{incomp}) are zero modes of ${\hat L}_1$. For the pressure term this can be obtained by direct differentiation:  
\begin{eqnarray}&&\!\!\!\!\!\!\!\!\! 
\frac{1}{\sqrt{\xi^2\!+\!\eta^2}}\left(3+\rho\frac{\partial}{\partial\rho}+z\frac{\partial}{\partial z}\right)P
=\int_0^{\infty} \!\!\left[(kB'+3B)\cosh(k\xi)\!+\!(kC'+3C)\sinh(k\xi)\right]J_{1}(k\eta) dk,
\end{eqnarray}
which satisfies the general form of a zero mode of ${\hat L}_1$. Above we used integration by parts and 
\begin{eqnarray}&&\!\!\!\!\!\!\!\!\! 
\eta\frac{\partial}{\partial\eta} J_{1}(k\eta)=k\frac{\partial}{\partial k}J_{1}(k\eta),\ \ 
\sqrt{\xi^2+\eta^2}=\frac{2}{\sqrt{\rho^2+z^2}},\ \ \rho\frac{\partial}{\partial\rho}+z\frac{\partial}{\partial z}=-\eta\frac{\partial}{\partial\eta}-\xi \frac{\partial}{\partial \xi}.
\end{eqnarray}
Note that the last term in the RHS of Eq.~(\ref{incomp}) is also a zero mode of ${\hat L}_1$. Indeed, ${\hat L}_1\phi=0$ implies that ${\hat L}_1\partial_z\phi=\partial_z{\hat L}_1\phi=0$. Thus the derivative satisfies the general form of a zero mode of ${\hat L}_1$
\begin{eqnarray}&&\!\!\!\!\!\!\!\!\! 
\frac{1}{\sqrt{\xi^2\!+\!\eta^2}}\frac{\partial\phi}{\partial z}=\int_0^{\infty}\left[x(k)\cosh(k\xi)+y(k)\sinh(k\xi)\right]J_1(k\eta)dk, \label{cjf}
\end{eqnarray}
where $x$ and $y$ are some functions of $k$. Direct differentiation of $\phi$ in Eqs.~(\ref{oneil}) gives
\begin{eqnarray}&&\!\!\!\!\!\!\!\!\! 
\frac{1}{\sqrt{\xi^2\!+\!\eta^2}}\frac{\partial\phi}{\partial z}
=\!-\frac{\xi}{2}\!\int_0^{\infty}\!\! A(k)\sinh(k\xi)J_{1}(k\eta)dk\!+\!\frac{\eta^2\!-\!\xi^2}{2}\!\int_0^{\infty}\!\! A(k)\cosh(k\xi)J_{1}(k\eta)k dk\!+\!\xi\!\int_0^{\infty}\!\!J_{1}(k\eta) dk\partial_k(k A\sinh(k\xi))
\nonumber\\&&\!\!\!\!\!\!\!\!\!
=\frac{\xi}{2}\int_0^{\infty}\!\! A(k)\sinh(k\xi)J_{1}(k\eta)dk+\frac{\xi^2}{2}\int_0^{\infty}\!\! A(k)\cosh(k\xi)J_{1}(k\eta)k dk+\xi\int_0^{\infty}\!\!A'\sinh(k\xi) J_{1}(k\eta) k dk-\frac{1}{2}\int_0^{\infty}\!\!\frac{dk}{k}\nonumber\\&&\!\!\!\!\!\!\!\!\!
\times A(k)\cosh(k\xi)\left(k^2\frac{d^2}{dk^2}\!+\!k\frac{d}{dk}\!-\!1\right)J_{1}(k\eta)
=\frac{\xi}{2}\int_0^{\infty}\!\! A(k)\sinh(k\xi)J_{1}(k\eta)dk+\frac{\xi^2}{2}\int_0^{\infty}\!\! A(k)\cosh(k\xi)J_{1}(k\eta)k dk \nonumber\\&&\!\!\!\!\!\!\!\!\!+\xi\int_0^{\infty}\!\!A'\sinh(k\xi) J_{1}(k\eta) k dk-\frac{1}{2}\int_0^{\infty}\!\!\Biggl[A''k\cosh(k\xi)\!+\!A'\cosh(k\xi)\!+\!2k\xi A'\sinh(k\xi)\!+\!\xi A\sinh(k\xi)\!+\!k\xi^2A\cosh(k\xi)
\Biggr.\nonumber\\&&\!\!\!\!\!\!\!\!\!\Biggl.
-\frac{A(k)\cosh(k\xi)}{k} \Biggr]J_{1}(k\eta)dk,\label{fd}
\end{eqnarray}
where we used 
\begin{eqnarray}&&\!\!\!\!\!\!\!\!\! 
\frac{\partial }{\partial z}\sqrt{\xi^2\!+\!\eta^2}=-\frac{\xi\sqrt{\xi^2+\eta^2}}{2};\ \ 
k^2\frac{d^2}{dk^2}J_1(k\eta)+k\frac{d}{dk} J_1(k\eta)-J_{1}(k\eta)=-k^2\eta^2J_{1}(k\eta).
\end{eqnarray} 
While the correspondence between Eqs.~(\ref{fd}) and (\ref{cjf}) is not straightforward, we can determine $f(k)$ and $g(k)$ from Eq.~(\ref{fd}) upon substituting $\xi=0$:  
\begin{eqnarray}&&\!\!\!\!\!\!\!\!\! 
\left.\left(\frac{1}{\sqrt{\xi^2\!+\!\eta^2}}\frac{\partial\phi}{\partial z}\right)\right|_{\xi=0}=-\frac{1}{2}\int_0^{\infty}\!\!\left(A''k+A'
-\frac{A(k)}{k} \right)J_{1}(k\eta)dk,\ \quad \left.\left(\frac{\partial}{\partial \xi}\frac{1}{\sqrt{\xi^2\!+\!\eta^2}}\frac{\partial\phi}{\partial z}\right)\right|_{\xi=0}=0.
\end{eqnarray}
Comparison with the corresponding expressions in Eq.~(\ref{cjf}) yields 
\begin{eqnarray}&&\!\!\!\!\!\!\!\!\! 
x(k)=-\frac{1}{2}\left(A''k+A'-\frac{A(k)}{k}\right),\ \ y(k)=0;\ \ 
\frac{1}{\sqrt{\xi^2\!+\!\eta^2}}\frac{\partial \phi}{\partial z}=-\frac{1}{2}\int_0^{\infty} \left(A''k+A'-\frac{A(k)}{k}\right)\cosh(k\xi)J_1(k\eta)dk.
\end{eqnarray}
Similar analysis determines the rest of the terms in the RHS of the continuity equation (\ref{incomp}). We have from Eq.~(\ref{lk})
\begin{eqnarray}&&\!\!\!\!\!\!\!\!\!
0=\partial_{\rho}{\hat L}_0 \psi={\hat L}_0 \partial_{\rho}\psi-\frac{1}{\rho^2}\partial_{\rho}\psi={\hat L}_1 \partial_{\rho}\psi,
\end{eqnarray}
so that $\partial_{\rho}\psi$ is a zero mode of ${\hat L}_1$. This implies that
\begin{eqnarray}&&\!\!\!\!\!\!\!\!\!\frac{1}{\sqrt{\xi^2\!+\!\eta^2}}\partial_{\rho}\psi
=\int_0^{\infty}\left(x_1(k)\cosh(k\xi)+y_1(k)\sinh(k\xi)\right)J_1(k\eta)dk,\label{x1}
\end{eqnarray}
with some functions $x_1(k)$ and $y_1(k)$. We have by differentiation
\begin{eqnarray}&&\!\!\!\!\!\!\!\!\!
\frac{\partial_{\rho}\psi}{\sqrt{\xi^2\!+\!\eta^2}}
\!=\!-\frac{\eta}{2}\int_0^{\infty} \!\!\left(D(k)\cosh(k\xi)\!+\!E(k)\sinh(k\xi)\right)J_{0}(k\eta) dk
\!-\!\frac{\xi^2\!-\!\eta^2}{2}
\int_0^{\infty} \!\!\left(D(k)\cosh(k\xi)\!+\!E(k)\sinh(k\xi)\right)J_1(k\eta) k dk
\nonumber\\&&\!\!\!\!\!\!\!\!\! 
-\eta \xi\int_0^{\infty} \!\!\left(D(k)\sinh(k\xi)\!+\!E(k)\cosh(k\xi)\right)J_{0}(k\eta) kdk,\nonumber
\end{eqnarray}
where we used that $2\partial_{\rho}\sqrt{\xi^2\!+\!\eta^2}=-\eta\sqrt{\xi^2+\eta^2}$.

It can further be shown that 
\begin{eqnarray}&&\!\!\!\!\!\!\!\!\!
\left.\frac{\partial_{\rho}\psi}{\sqrt{\xi^2\!+\!\eta^2}}\right|_{\xi=0}
=\!\frac{1}{2}\int_0^{\infty} \!\!\left(\frac{D}{k}\right)'J_1(k\eta) k dk\!-\!\frac{1}{2}
\int_0^{\infty} \!\!J_1(k\eta)\left(k^2\frac{d^2}{dk^2}\!-\!k\frac{d}{dk}\!-\!1\right)D(k) \frac{dk}{k}
\!=\!-\frac{1}{2}
\int_0^{\infty} \!\!J_1(k\eta)\left((kD)''\!-\!2D'\right) dk,\nonumber
\end{eqnarray}
whose comparison with Eq.~(\ref{x1}) gives $x_1=-kD''/2$.
Similarly, it is readily seen by comparison with the previous derivation that
\begin{eqnarray}&&\!\!\!\!\!\!\!\!\!
\left.\left(\frac{\partial}{\partial \xi}\frac{\partial_{\rho}\psi}{\sqrt{\xi^2\!+\!\eta^2}}\right)\right|_{\xi=0}
=\frac{3}{2}\int_0^{\infty} \!\!E'J_1(k\eta) k dk-\frac{1}{2}
\int_0^{\infty} \!\!J_1(k\eta)\left(k^2\frac{d^2}{dk^2}-k\frac{d}{dk}-1\right)E(k) dk
=-\frac{1}{2}
\int_0^{\infty} \!\!J_1(k\eta)k^2E''dk,
\end{eqnarray}
which implies $y_1=-kE''/2$. 
In the last two equations we used the following relations between the Bessel functions: 
\begin{eqnarray}&&\!\!\!\!\!\!\!\!\! 
J_0(k\eta)=\frac{1}{k\eta}
\frac{d(k J_1(k\eta))}{dk},\ \qquad J_2(k\eta)=\frac{J_1(k\eta)}{k\eta}-\frac{1}{\eta} \frac{\partial J_1(k\eta)}{\partial k}. \label{rc}
\end{eqnarray}
Thus conclude that
\begin{eqnarray}&&\!\!\!\!\!\!\!\!\!
\partial_{\rho}\psi
=-\frac{1}{2} \sqrt{\xi^2\!+\!\eta^2}\int_0^{\infty}\left[kD''\cosh(k\xi)+kE''\sinh(k\xi)\right] J_1(k\eta)dk,
\end{eqnarray}
Finally, we notice using ${\hat L}_2={\hat L}_1-3/\rho^2$ that 
\begin{eqnarray}&&\!\!\!\!\!\!\!\!\! 
0=\partial_{\rho}{\hat L}_2\chi=
{\hat L}_2\partial_{\rho}\chi-\frac{1}{\rho^2}\partial_{\rho}\chi+\frac{8}{\rho^3}\chi,\ \ 
{\hat L}_1\partial_{\rho}\chi=\frac{4}{\rho^2}\partial_{\rho}\chi-\frac{8}{\rho^3}\chi,
\end{eqnarray}
We also notice that 
\begin{eqnarray}&&\!\!\!\!\!\!\!\!\! 
{\hat L}_1\frac{1}{\rho}\chi=\frac{1}{\rho}{\hat L}_1\chi-\frac{2}{\rho^2}\frac{\partial \chi}{\partial \rho}
+\frac{1}{\rho^3}
\chi=-\frac{2}{\rho^2}\frac{\partial \chi}{\partial \rho}
+\frac{4}{\rho^3}\chi.
\end{eqnarray}
Summation of the last two equations shows that ${\hat L}_1\left(\partial_{\rho}\chi+2\chi/\rho\right)=0$. Therefore we have  
\begin{eqnarray}&&\!\!\!\!\!\!\!\!\! 
\frac{1}{\sqrt{\xi^2\!+\!\eta^2}}\left(\partial_{\rho}\chi+\frac{2\chi}{\rho}\right)=\int_0^{\infty}\left[x_2(k)\cosh(k\xi)+y_2(k)\sinh(k\xi)\right] J_1(k\eta)dk,\label{tldx}
\end{eqnarray}
for some functions $x_2(k)$ and $y_2(k)$. Differentiation of the last of Eqs.~(\ref{oneil}) yields 
\begin{eqnarray}&&\!\!\!\!\!\!\!\!\! 
\frac{\partial_{\rho}\chi\!+\!2\chi/\rho}{\sqrt{\xi^2\!+\!\eta^2}}\!=\!\frac{\eta}{2}\!\int_0^{\infty} \!\!\left[F(k)\cosh(k\xi)\!+\!G(k)\sinh(k\xi)\right] J_{2}(k\eta) dk
\!+\!\frac{\xi^2}{\eta}\!\int_0^{\infty} \!\!\left[F(k)\cosh(k\xi)\!+\!G(k)\sinh(k\xi)\right]J_{2}(k\eta) dk
\nonumber\\&& \!\!\!\!\!\!\!\!\!
+\frac{\xi^2-\eta^2}{2}\int_0^{\infty} \!\!\left[F(k)\cosh(k\xi)\!+\!G(k)\sinh(k\xi)\right] J_{2}'(k\eta) k dk-\eta\xi\int_0^{\infty} \!\!\left[F(k)\sinh(k\xi)\!+\!G(k)\cosh(k\xi)\right]J_{2}(k\eta)kdk. \label{LV}
\end{eqnarray}
By setting $\xi=0$ in the last equation we obtain 
\begin{eqnarray}&&\!\!\!\!\!\!\!\!\! 
\left.\frac{\partial_{\rho}\chi\!+\!2\chi/\rho}{\sqrt{\xi^2\!+\!\eta^2}}\right|_{\xi=0}
=\frac{\eta}{2}\int_0^{\infty} \!\!F(k)J_{2}(k\eta) dk
-\frac{\eta^2}{2}\int_0^{\infty} \!\!F(k)J_{2}'(k\eta) k dk=\frac{\eta}{2}\int_0^{\infty} \!\!(kF'+2F)J_{2}(k\eta) dk
\nonumber\\&&\!\!\!\!\!\!\!\!\!
=\frac{1}{2}\int_0^{\infty} \!\!(F'+2F/k)J_{1}(k\eta) dk+
\frac{1}{2}\int_0^{\infty} \!\!(kF'+2F)'J_{1}(k\eta) dk, \nonumber 
\end{eqnarray}
where we used the 2nd of Eqs.~(\ref{rc}). The comparison of the last result with Eq.~(\ref{tldx}) readily gives $x_2(k)= kF''/2+2 F'+F/k$. Analogously we have
\begin{eqnarray}&&\!\!\!\!\!\!\!\!\! \left.\left(\frac{\partial}{\partial \xi}\frac{\partial_{\rho}\chi\!+\!2\chi/\rho}{\sqrt{\xi^2\!+\!\eta^2}}\right)\right|_{\xi=0}\!\!=\!\frac{\eta}{2}\int_0^{\infty} \!\!(kG'+G)J_{2}(k\eta)k dk=
\frac{1}{2}\int_0^{\infty} \!\!(kG'+G)J_{1}(k\eta)dk+
\frac{1}{2}\int_0^{\infty} \!\!(k^2G'+kG)'J_{1}(k\eta) dk.
\end{eqnarray}
which gives ${y_2}=kG''/2+2G'+G/k$ such that
\begin{eqnarray}&&\!\!\!\!\!\!\!\!\! 
\partial_{\rho}\chi+\frac{2\chi}{\rho}\!=\!\frac{\sqrt{\xi^2\!+\!\eta^2}}{2}\int_0^{\infty}\! \left[\left(kF''\!+\!4F'\!+\!\frac{2F}{k}\right)\cosh(k\xi)\!+\!\left(kG''\!+\!4G'\!+\!\frac{2G}{k}\right)\sinh(k\xi)\right] J_1(k\eta)dk.\nonumber
\end{eqnarray}
Finally, combining the above results the continutity equation yields the following ODE's: 
\begin{eqnarray}&&\!\!\!\!\!\!\!\!\! 
2\left(k\frac{d}{dk}+3\right)\left( \begin{array}{c}
B \\
C \end{array} \right)-2\left(k\frac{d^2}{dk^2}+\frac{d}{dk}-\frac{1}{k}\right)\left( \begin{array}{c}
A \\
0 \end{array} \right)-k\frac{d^2}{dk^2}\left( \begin{array}{c}
D \\
E \end{array} \right)+\left(k\frac{d^2}{dk^2}+4\frac{d}{dk}+\frac{2}{k}\right)\left( \begin{array}{c}
F \\
G \end{array} \right)=0. \label{ode1}
\end{eqnarray}
Ref.~\cite{nir1973} considered a more general case of a motion of a dumbbell made of two touching spheres in a linear shear flow. Their
Eqs.~2.9 (for $m=1$) are identical to Eqs.~(\ref{ode1}) except for a factor of $2$ in the first term in the LHS.

\subsection{Boundary conditions}

In order to determine the seven unknown functions in the solution given by Eqs.~(\ref{oneil}), besides the continuity equation, we shall use the no-slip boundary conditions at the plane ($\xi=0$), and at the sphere surface ($\xi=1$). Below we follow \cite{o1968sphere} and express the six functions in terms of a single function $A(k)$. Then the system of Eqs.~(\ref{ode1}) transforms into a single ODE for $A(k)$ that can be solved numerically.

\begin{figure}
    \includegraphics[width=0.45\textwidth]{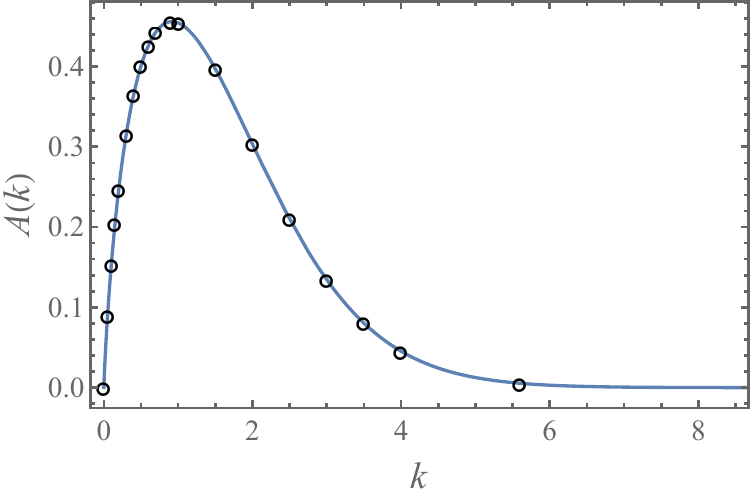}
    \caption{
        \label{fig:Ak}
        The plot of $A$ vs. $k$ (solid line) obtained by the numerical integration of Eq.~(\ref{ode2}). The symbols are the data from Table~1 in Ref.~\cite{o1968sphere}.  
    }
\end{figure}

We begin with the boundary condition on the vertical velocity $W$. The condition $W=0$ at $\xi=0$ was already applied. However, the continuity equation (\ref{incomp}) together with $U=V=\partial_\rho U=0$ at $\xi=0$ provides an extra condition, $\partial_z W(\xi=0)=0$, which results in 
\begin{eqnarray}&&\!\!\!\!\!\!\!\!\! 
\int_0^{\infty} \!\! B(k)J_{1}(k\eta) dk
\!=\!-\eta^2\int_0^{\infty}\!\! A(k)J_{1}(k\eta) k dk\!=\!\int_0^{\infty}\!\! A(k)\left(k^2\frac{d^2}{dk^2}J_1(k\eta)+k\frac{d}{dk} J_1'(k\eta)-J_{1}(k\eta)\right)  \frac{dk}{k}.\label{fs}
\end{eqnarray}
where we used the equation for the Bessel function of order one $z^2J_1''+z J_1'-J_1=-z^2J_1$. Integrating by parts the last term in Eq.~(\ref{fs}) and equating the coefficients of the Bessel function on both sides, we obtain:  
\begin{eqnarray}&&\!\!\!\!\!\!\!\!\! 
B\!=\!(kA)''-A'-\frac{A}{k}=kA''+A'-\frac{A}{k}. \label{ba}
\end{eqnarray}
The condition at the particle surface  $W(\xi=1)=0$ gives 
\begin{eqnarray}&&\!\!\!\!\!\!\!\!\!
-\frac{1}{1\!+\!\eta^2}
\int_0^{\infty} \!\!\left(B(k)\cosh k\!+\!C(k)\sinh k\right)J_{1}(k\eta) dk=
\int_0^{\infty} \!\! A(k)\sinh k J_{1}(k\eta) dk, \nonumber
\end{eqnarray}
which can be rewritten in the form similar to that in Eq.~(\ref{fs}):
\begin{eqnarray}&&\!\!\!\!\!\!\!\!\!
\int_0^{\infty} \!\!\left[B\cosh k\!+\!(C+A)\sinh k\right] J_{1}(k\eta) dk=
-\eta^2 \int_0^{\infty} \!\! A(k)\sinh k J_{1}(k\eta) dk, \nonumber
\end{eqnarray}
which gives the corresponding condition
\begin{eqnarray}&&\!\!\!\!\!\!\!\!\!
B\cosh k\!+\!(C\!+\!A)\sinh k\!=  k \left(\frac{A\sinh k}{k}\right)''\!+\!\left(\frac{A\sinh k}{k}\right)'\!-\!\frac{A\sinh k}{k^2}\!=\! \nonumber \\
&& A''\sinh k\!-\!\frac{A'\sinh k}{k}\!+\!2A'\cosh k
\!+\!A\sinh k -\frac{A \cosh k}{k}. \nonumber
\end{eqnarray}
Using the Eq.~(\ref{ba}) for $B$ the function $C$ can be readily obtained as
\begin{eqnarray}&&\!\!\!\!\!\!\!\!\!
C\sinh k=
A''\sinh k-k\cosh k A''-\frac{A'\sinh k}{k}+A'\cosh k \  \Rightarrow \  C=\left(kA''-A'\right)K,\nonumber
\end{eqnarray}
where following Ref.~\cite{o1968sphere} we introduced:
\[
\ K\equiv k^{-1}-\coth k\,.
\]
The conditions of vanishing of $U$ and $V$ at $\xi=0$ give 
\begin{eqnarray}&&\!\!\!\!\!\!\!\!\! 
\int_0^{\infty} \!\! B(k)J_{1}(k\eta) dk\!=\!-\eta\int_0^{\infty} \!\!D(k)J_{0}(k\eta) dk,\ \quad  \int_0^{\infty} \!\! B(k)J_{1}(k\eta) dk\!=\!-\eta\int_0^{\infty} \!\!F(k)J_{2}(k\eta) dk, \nonumber
\end{eqnarray}
where we used Eq.~(\ref{oneil}). Using in the above equations the relations in Eqs.~(\ref{rc}) and integration by parts, we find that $B=k(D/k)'$ and  $kB=-(kF)'$. This gives expressions for $D$ and $F$ via $A$ by using Eq.~(\ref{ba}):
\begin{eqnarray}
\frac{d}{dk}\frac{D}{k}=\frac{d}{dk} \left(A'+\frac{A}{k}\right) \ \ & \Rightarrow &\ \ \ D=kA'+A,\nonumber \\ 
-\frac{d(kF)}{dk}=\frac{d}{dk} (k^2A'-kA) \ \ & \Rightarrow & \ \  F=A-kA', \nonumber
\end{eqnarray}
where we made use of the fact that the coefficients vanish as $k\to\infty$. The remaining conditions are $U=-V=-z$ at the particle surface $\xi=1$, see Eq.~(\ref{a}), resulting in:
\begin{eqnarray}&&\!\!\!\!\!\!\!\!\! 
\eta\int_0^{\infty}\!\! A(k)\sinh k J_{1}(k\eta)dk
-\frac{4}{(1+\eta^2)^{3/2}}\!=\!\int_0^{\infty} \!\!\left[D(k)\cosh k\!+\!E(k)\sinh k\right] J_{0}(k\eta) dk,\nonumber\\&&\!\!\!\!\!\!\!\!\! \eta\int_0^{\infty}\!\! A(k)\sinh k J_{1}(k\eta)dk\!=\!\int_0^{\infty} \!\!\left[F(k)\cosh k\!+\!G(k)\sinh k\right] J_{2}(k\eta) dk.\label{ks}
\end{eqnarray}
For the first in Eqs.~(\ref{ks}) we have
\begin{eqnarray}&&\!\!\!\!\!\!\!\!\! 
\eta\int_0^{\infty}\!\! A(k)\sinh k J_{1}(k\eta)dk=-\int_0^{\infty}\!\! A(k)\sinh k \frac{d}{dk}J_0(k\eta)dk=\int_0^{\infty}\!\! (A\sinh k)'J_0(k\eta)dk,\nonumber\\&&
\!\!\!\!\!\!\!\!\! 
\frac{1}{(1+\eta^2)^{3/2}}=\int_0^{\infty} ke^{-k} J_0(k\eta)dk,
\end{eqnarray}
where we assumed that $A(k)\sinh k$ vanishes at $k=0$. We thus find 
\begin{eqnarray}&&\!\!\!\!\!\!\!\!\! 
(A\sinh k)'- 4k e^{-k}=D(k)\cosh k\!+\!E(k)\sinh k,
\end{eqnarray}
which upon substituting $D=kA'+A$ becomes 
\begin{eqnarray}&&\!\!\!\!\!\!\!\!\! 
E= kA'K- 4k(\coth k-1). \nonumber
\end{eqnarray}
Finally, using $k\eta J_0(k\eta)+k\eta J_2(k\eta)=2J_1(k\eta)$ in the 2nd of Eqs.~(\ref{ks}) gives
\begin{eqnarray}&&\!\!\!\!\!\!\!\!\! 
-\eta^2\!\int_0^{\infty}\!\! A(k)\sinh k J_{1}(k\eta)dk\!=\!\eta\!\int_0^{\infty} \!\!\left[F(k)\cosh k\!+\!G(k)\sinh k\right] J_0(k\eta) dk\!-\!2\!\int_0^{\infty} \!\!\left[F(k)\cosh k\!+\!G(k)\sinh k\right] J_{1}(k\eta) \frac{dk}{k}.\nonumber
\end{eqnarray}
This can be rewritten as
\begin{eqnarray}&&\!\!\!\!\!\!\!\!\! 
\int_0^{\infty}\!\! A(k)\left(k^2\frac{d^2}{dk^2}J_1(k\eta)+k\frac{d}{dk} J_1'(k\eta)-J_{1}(k\eta)\right)  \frac{\sinh k dk}{k^2}
\nonumber\\&&\!\!\!\!\!\!\!\!\!
=\int_0^{\infty} \!\!\left(F(k)\cosh k\!+\!G(k)\sinh k\right) \frac{d(k J_1(k\eta))}{dk} \frac{dk}{k}-2\int_0^{\infty} \!\!\left(F(k)\cosh k\!+\!G(k)\sinh k\right)J_{1}(k\eta) \frac{dk}{k}.
\end{eqnarray}
where Eq.~(\ref{rc}) was used. Integrating by parts and comparing the two sides of the equation, we find   
\begin{eqnarray}&&\!\!\!\!\!\!\!\!\! 
(A\sinh k)''-\left(\frac{A\sinh k}{k}\right)'-\frac{A\sinh k}{k^2}
=-(F\cosh k\!+\!G\sinh k)'-\frac{F\cosh k\!+\!G\sinh k}{k}. \nonumber
\end{eqnarray}
Substituting $F=A-kA'$, we find after some algebra
\begin{eqnarray}&&\!\!\!\!\!\!\!\!\! 
kKA''-(k+K)A'+2A= -G'-G\coth k -\frac{G(k)}{k}\ \ \Rightarrow \ \ G=\left(2A-kA'\right)K.
\end{eqnarray}
Collecting all coefficients in terms of $A$:
\begin{eqnarray}&&\!\!\!\!\!\!\!\!\! 
B\!=\!kA''+A'-\frac{A}{k},\ \ C=\left(kA''-A'\right)K,\ \ D=kA'+A, \nonumber\\&&\!\!\!\!\!\!\!\!\! 
E= kA'K- 4k(\coth k-1), \ \ F=A-kA', \ \  G=\left(2A-kA'\right)K.
\label{allcoef}
\end{eqnarray}
Substituting the above relationship between the coefficients into Eq.~(\ref{ode1}) we obtain two 2$^\mathrm{nd}$-order ODE's for $A(k)$:
\begin{eqnarray}&&\!\!\!\!\!\!\!\!\! 
0=2\left( \begin{array}{c}
k^2A'''+5kA''+2A'-2A/k\\
k\left(kA''-A'\right)K'+k^2 A'''K+3\left(kA''-A'\right)K \end{array} \right)-2\left( \begin{array}{c}
kA''+A'-A/k \\
0 \end{array} \right)
\label{ros}\nonumber\\&&\!\!\!\!\!\!\!\!\!
-\left( \begin{array}{c}
k^2A'''+3kA'' \\
k^2A'''K+kA'(kK)''+2kA''(kK)'-4k^2(\coth k-1)''-8k(\coth k-1)' \end{array} \right)
\nonumber\\&&\!\!\!\!\!\!\!\!\!
+\left( \begin{array}{c}
-k^2A'''-5kA''-2A'+2A/k \\
-k^2A'''K+2k\left(A'-kA''\right)K'
+k\left(2A-kA'\right)K'' +
4\left(A'-kA''\right)K+4\left(2A-kA'\right)K'+2\left(2A-kA'\right)K/k \end{array} \right)\nonumber.
\end{eqnarray}
The equation in the second row is satisfied trivially due to  the requirement $\partial_z W=0$ at $\xi=0$, which was already incorporated into the solution. The first row of Eqs.~(\ref{ros}) yields 
\begin{eqnarray}&&\!\!\!\!\!\!\!\!\! 
k^3 K' A'' +kA'\left(k^2K''+2K+3kK'\right)-A\left(k^2  K'' +4k K'+2K\right)=2k^2(2X'+kX'')\,, \label{ode2}
\end{eqnarray}
where $X\equiv \coth k-1$. This equation differs from the corresponding Eq.~$3.19$ in \cite{o1968sphere}, where the coefficient $k K'$ of $A''$ on the LHS should be replaced by $k^3 K'$ (notice that Ref.~\cite{o1968sphere} uses the variable $s$ instead of $k$ \cite{note1}). The coefficient $k^3 K'$ is also consistent with small-$k$ asymptotic form, i.e., the Euler equation, giving the particular solution, $A_\mathrm{I}\approx 4k$ and the two linearly independent homogeneous solutions $A_\mathrm{II,III}=k^{\pm\sqrt{10}-2}$, provided in \cite{o1968sphere} following Eq.~3.19. 
The function $A(k)$ is determined by solving the Eq.~(\ref{ode2}) numerically by \textit{Mathematica} as the two-point boundary value problem  on the interval $[\epsilon, L]$, where $\epsilon=10^{-4}$ and $L=15$, with the following boundary conditions: $A(L)=0$ and $A(\epsilon)=4\epsilon$. The later condition follows from the asymptotic small-$k$ expansion of the solution of Eq.~(\ref{ode2})  given by $A\simeq 4 k + C_1 k^{-2 - \sqrt{10}} + C_2 k^{-2 + \sqrt{10}}$, where $C_1=0$ to avoid divergence. Fig.~\ref{fig:Ak} depicts the numerical solution for $A(k)$ (solid line), while the symbols stand for the data from Table $1$ of \cite{o1968sphere}, showing an excellent agreement, indicating that the incorrect coefficient in Eq.~$3.19$ in \cite{o1968sphere} is actually a typo.  

\begin{figure}
    \includegraphics[width=0.36\textwidth]{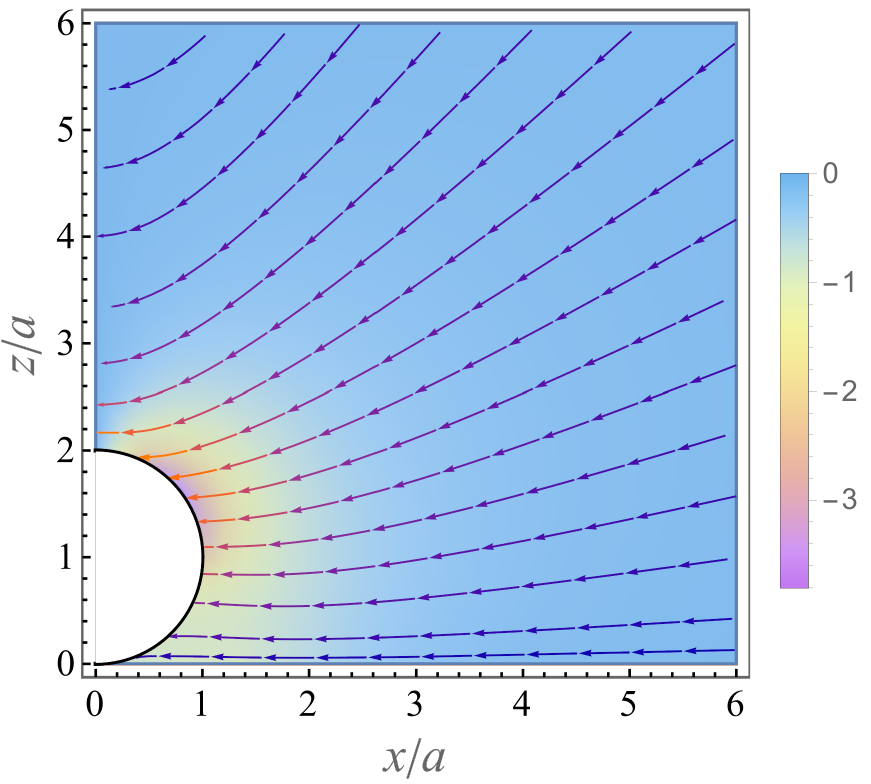}
    \caption{
        \label{fig:flow}
        The perturbation of the simple shear flow, $\bm v=\bm w\!-\!z\hat{\bm x}$ (streamlines) and pressure $p$ (color map) in Eqs.~(\ref{ca}) due to a spherical particle in contact with the plane wall at $z\!=\!0$ in $xz$-plane (for $\varphi\!=\!0$). All lengths are scaled with the particle radius $a$, velocity with $a\dot{\gamma}$ and pressure with $\eta\dot{\gamma}$.  
    }
\end{figure}

Computing the rest of the coefficients in Eqs.~(\ref{allcoef}) via $A(k)$ and using the solutions for scalar harmonic functions (\ref{oneil}) and the ansatz (\ref{ca}), one can compute and depict the flow disturbance $\bm v$ and the pressure $p$ around the particle (see Fig.~\ref{fig:flow}). It can be readily observed that the presence of the particle hinders the shear flow. The pressure at the particle surface facing the impinging shear flow increases, while it decreases at the rear, such that the flow exerts the viscous force $F_x>0$ and torque $L_y>0$.       

\subsection{Viscous force and torque exerted on the particle}

It can be readily shown that the undisturbed linear shear flow $\bm w\!=\!z\hat{\bm x}$ does not contribute to the force or the torque exerted on the attached particle, therefore they can be determined from the perturbed flow in Eqs.~(\ref{eqv}). Below we present an alternative and simpler derivation to that of O'Neill \cite{o1968sphere} (see Appendix~\ref{app:A}) and correct some typos showing in his paper. 

The force that the flow exerts on the particle reads
\begin{equation}
F_x =\oint_{r=1}\sigma_{xk} dS_k\,.\label{force}    
\end{equation}
It is useful to rewrite Eq.~(\ref{eqv}) in the form $\nabla_k\sigma_{ik}=0$ which gives upon the integration over the whole space:
\begin{equation}
F_i\equiv \oint_{r=1}\sigma_{ik} dS_k= -\int_{z=0} \sigma_{iz} dxdy\,, 
\end{equation}
where we used the fact that the integral over the infinite semi-sphere vanishes (see Sec.~\ref{sec:farfield}). Therefore,
\begin{equation}
F_x=-\int_{z=0} \frac{\partial v_x}{\partial z}dxdy=-\int_{z=0} \left(\cos^2\theta \frac{\partial U(\rho, z)}{\partial z}-\sin^2\theta \frac{\partial V(\rho, z)}{\partial z}\right)\rho d\rho d\theta\,,
\end{equation}
where we used Eqs.~(\ref{ca}) and (\ref{angl}). Thus we find 
\begin{eqnarray}&&\!\!\!\!\!\!\!\!\! 
F_x=-\pi\int_{z=0} 
\left(\frac{\partial \psi}{\partial z} +\frac{\rho}{2}\frac{\partial P}{\partial z}\right) 
\rho d\rho=-2\pi\int_0^{\infty}\left. \left(
\eta\frac{\partial \psi}{\partial \xi} +\frac{\partial P}{\partial \xi} \right)\right|_{\xi=0}\; \frac{d\eta}{\eta^2}\,,
\end{eqnarray}
where we used $\rho(\xi=0)=2/\eta$. It follows from Eqs.~(\ref{oneil}) that
\begin{eqnarray}&&\!\!\!\!\!\!\!\!\! 
\left.\frac{\partial P}{\partial \xi}\right|_{\xi=0}\!=\!\eta\int_0^{\infty} C(k)J_{1}(k\eta) kdk;\ \quad \left.\frac{\partial \psi
}{\partial \xi}\right|_{\xi=0}\!=\!\eta\int_0^{\infty}E(k)J_{0}(k\eta) k dk.
\end{eqnarray}
We therefore conclude that 
\begin{eqnarray}&&\!\!\!\!\!\!\!\!\! 
F_x=-2\pi\int_0^{\infty} \left(\int_0^{\infty}E(k)J_{0}(k\eta) k dk+\frac{1}{\eta}\int_0^{\infty} C(k)J_{1}(k\eta) kdk\right) d\eta=2\pi\int_0^{\infty} \frac{d\eta}{\eta}\int_0^{\infty}(E'-C)J_1(k\eta)  kdk \nonumber\\&&\!\!\!\!\!\!\!\!\!
=2\pi\int_0^{\infty}(E'-C)kdk=2\pi\int_0^{\infty}\left\{kA'K'+2KA'- 4(k(\coth k-1))'\right\}kdk 
\nonumber\\&&\!\!\!\!\!\!\!\!\!
=2\pi\int_0^{\infty}\left\{4k(\coth k-1)-k^2KA''\right\}dk 
=2\pi\int_0^{\infty}\left\{4k(\coth k-1)+(k^2\coth k)''A\right\}dk, \label{Fx}
\end{eqnarray}
which agrees with \cite{o1968sphere}. Using the numerical solution for $A(k)$ in  Fig.~\ref{fig:Ak} and in (\ref{Fx}) yields upon integration 
\[
F_x=6\pi f\,,\quad f\simeq 1.70049,\,
\]
which is close to $f\simeq 1.7009$ in \cite{o1968sphere}. The extrapolated (to a vanishing separation distance from the plane) result corresponding to a freely suspended particle in a simple shear flow by in Ref.~\cite{Goldman1967b} gives practically the same result, $f\simeq 1.7005$. The other  components of the force are zero. Clearly, the force on the particle in the the direction of the shear flow. 

Next we consider the torque $\bm L$ that the shear flow applies on the particle attached to the plane. The Stokes equations imply that $\nabla_k \epsilon_{lji}\left(\bm r-\bm r_0\right)_j\sigma_{ik}=0$ holds for any $\bm r_0$. Integrating this identity over the flow domain gives (the integral over the infinite semi-sphere vanishes, see Sec.~\ref{sec:farfield})
\begin{eqnarray}&&\!\!\!\!\!\!\!\!\! 
\oint_{r=1}\epsilon_{lji}\left(\bm r-\bm r_0\right)_j\sigma_{ik}d{\hat n}_k=-\int_{z=0}\epsilon_{lji}\left(\bm r-\bm r_0\right)_j\sigma_{iz}dxdy=-\int_{z=0}\epsilon_{lji}r_j\sigma_{iz}dxdy
+\int_{z=0}\epsilon_{lzi}\sigma_{iz}dxdy,
\end{eqnarray}
where in the last equality we took $\bm r_0$ at the center of the sphere, i.e., $\bm r_0=(0, 0, 1)$ and no summation is assumed over $z$. It can be readily shown that the $x$ component of the torque vanishes: 
\begin{eqnarray}&&\!\!\!\!\!\!\!\!\! 
L_x=-\int_{z=0}\sigma_{yz}dxdy-\int_{z=0}\epsilon_{xik}r_i\sigma_{kz}dxdy=-\int_{z=0}\frac{\partial v_y}{\partial z}dxdy-\int_{z=0}y\sigma_{zz}dxdy=\int_{z=0}ypdxdy=0\,.
\end{eqnarray}
The non-vanishing component is in $y-$direction. We have 
\begin{eqnarray}&&\!\!\!\!\!\!\!\!\! 
L_y+F_x=-\int_{z=0}\epsilon_{yik}r_i\sigma_{kz}dxdy=\int_{z=0}x\sigma_{zz}dxdy=-\int_{z=0}xpdxdy=-
\pi \int_{0}^{\infty}\rho^2 P(\rho, z=0)\, d\rho,
\end{eqnarray}
where we used $\int_{z=0}\epsilon_{yzi}\sigma_{iz}dxdy=\int_{z=0}\sigma_{xz}dxdy=-F_x$ and the arbitrary constant in pressure is fixed by the demand of pressure vanishing at infinity. Using Eq.~(\ref{oneil}), where $z=0$ is $\xi=0$, we find that 
\begin{eqnarray}&&\!\!\!\!\!\!\!\!\! 
L_y+F_x=-\pi \int_{0}^{\infty}\frac{8 d\eta}{\eta^3}
\int_0^{\infty} B(k)J_{1}(k\eta) dk=\pi \int_{0}^{\infty}\frac{8 d\eta}{\eta^3}
\int_0^{\infty} (A'+A/k)\frac{d}{dk}kJ_{1}(k\eta) dk
\nonumber\\&&\!\!\!\!\!\!\!\!\! 
=\pi \int_{0}^{\infty}\frac{8 d\eta}{\eta^2}
\int_0^{\infty} (kA'+A)J_{0}(k\eta) dk
=\pi \int_{0}^{\infty}\frac{8 d\eta}{\eta}
\int_0^{\infty} kAJ_1(k\eta) dk=8\pi \int_0^{\infty} kA dk, \label{Ly}
\end{eqnarray}
where we used $B(k)$ from Eq.~(\ref{allcoef}) re-written as $B\!=\!k(A'+A/k)'$. Numerical integration of the solution in Eq.~(\ref{Ly}) gives
\begin{equation}
L_y = -F_x +8\pi \int_0^{\infty} kA\, dk = 4\pi g,\quad g\simeq 0.94411\,. \label{Ly}
\end{equation}
The vaue of $g$ close to that reported in \cite{o1968sphere}, $g\simeq 0.943993$, however some typos are involved. The torque in  Ref.~\cite{o1968sphere} is defined as $L_y=-8\pi g$, which differs from our result by a factor of $2$ and the negative sign. Notice that the torque should be positive, $L_y>0$, as the shear flow acts to rotate the particle in the clock-wise direction about the $y$-axis. Besides the value of $g$, the torque in the Eq.~4.9 of Ref.~\cite{o1968sphere} is showing as $L_y=F_x+8\pi \int_0^{\infty} kA dk$, where the sign of the first term is wrong (c.f. Eq.~\ref{Ly}). 

This concludes the revised analysis of the problem posed in \cite{o1968sphere} and provides a detailed and self-consistent derivation free of typos. The solution for the flow is given in special ``tangent sphere" coordinates and it depends on a function $A(k)$ that can be determined numerically (see Fig.~\ref{fig:Ak}).  Therefore it is worth to derive an asymptotic limit of the flow far from the plane where it may simplify considerably. 

\section{Integral representation and far-field flow \label{sec:farfield}}

In this Section we introduce the integral representation of the solution and derive the far-field form of the flow disturbance. This analysis holds for adsorbed particles of an \textit{arbitrary} shape $S$ (it is still assumed that the origin is the point of contact with the plane) and demonstrates universal features of the far-flow. The case of surface corrugations will be considered elsewhere. 

The integral representation of the flow is obtained from the Lorentz identity 
\begin{eqnarray}&&\!\!\!\!\!\!\!\!\! 
\nabla_l (v_i\sigma_{il}^k)=\nabla_l (u_i^k\sigma_{il})-v_k\delta(\bm x-\bm x'),\label{loret}
\end{eqnarray}
where $\bm u^k$ is the flow due to Stokeslet near an infinite (no-slip) plane,
\begin{eqnarray}&&\!\!\!\!\!\!\!\!\! 
-\nabla_i p^k+\nabla^2 u^k_i=-\delta_{ik}\delta(\bm x-\bm x'),\ \ \nabla\cdot \bm u^k=0,\ \ 
\bm u^k(z=0)=\bm u^k(x=\infty)=0,
\end{eqnarray}
and $\sigma_{il}^k=-p^k\delta_{il}+\nabla_iu_l^k+\nabla_lu_i^k$ is the stress tensor associated with this flow, $\nabla_l\sigma_{il}^k=-\delta_{ik}\delta(\bm x-\bm x')$. Integration of Eq.~(\ref{loret}) over the whole volume of the flow $\bm v$ gives 
\begin{eqnarray}&&\!\!\!\!\!\!\!\!\! 
\int_{S}v_i\sigma_{il}^k dS_l=\int_{S} u_i^k\sigma_{il}dS_l +v_k(\bm x'), \label{sd}
\end{eqnarray}
where we assumed vanishing of the contributions of integrals over the surface at infinity (see below). Infinitesimal vector surface element $d\bm S$ is assumed to point into the fluid. For a sphere $\sigma_{il}^k dS_l=\sigma_{ir}^k dS$, while for now we use the general surface element. Renaming $\bm x\to\bm x'$, $i\to k$ and using $u^k_{i}(\bm x, \bm x')=u^i_{k}(\bm x', \bm x)$ we find the integral flow representation  
\begin{eqnarray}&&\!\!\!\!\!\!\!\!\! 
v_i(\bm x)
\!=\!\int_{S'}\! v_k(\bm x')\sigma_{kl}^i(\bm x', \bm x) dS_{l}'\!-\!\int_{S'} \!\! u^k_{i}(\bm x, \bm x')\sigma_{kl}(\bm x')dS_{l}'\!=\!-\int_{S'} \!\! z'\sigma_{xl}^i(\bm x', \bm x) dS_{l}'\!-\!\int_{S'}\! u^k_{i}(\bm x, \bm x')\sigma_{kl}(\bm x')dS_{l}', \label{rep}
\end{eqnarray}
where we also used the boundary condition on the sphere (for shear flow that condition is the same for any $S$). For the first integral in the RHS of the above equation we have 
\begin{eqnarray}&&\!\!\!\!\!\!\!\!\! 
\int_{S'} \! z'\sigma_{xl}^i(\bm x', \bm x) dS_{l}'=\int_{V'} \nabla'_l(z'\sigma_{xl}^i(\bm x', \bm x))dV'=\int_{V'} \sigma_{xz}^i(\bm x', \bm x)dV'=\int_{V'} \nabla'\cdot\left(u^i_x(\bm x', \bm x){\hat z}'+u^i_z(\bm x', \bm x){\hat x}'\right)dV'
\nonumber\\&&\!\!\!\!\!\!\!\!\! 
=\int_{S'} \left(u_i^x(\bm x, \bm x'){\hat z}'+u_i^z(\bm x, \bm x'){\hat x}'\right)\cdot d\bm S',\label{inf}
\end{eqnarray}
where $\nabla'$ is the derivative with respect to $\bm x'$ and $V'$ stands for the particle's interior. The above transformation allows to write the flow solely in terms of $u_i^{k}(\bm x, \bm x')$,
\begin{eqnarray}&&\!\!\!\!\!\!\!\!\! 
v_i(\bm x)
\!=\!-
\int_{S'} u_i^k(\bm x, \bm x')\left(\delta_{kx}\delta_{zl}+\delta_{kz}\delta_{xl}+\sigma_{kl}(\bm x')\right)dS_l'\equiv -\frac{1}{8\pi}\int_{S'} G_{ik}(\bm x, \bm x')\left(\delta_{kx}\delta_{zl}+\delta_{kz}\delta_{xl}+\sigma_{kl}(\bm x')\right)dS_l',\label{onl}
\end{eqnarray}
where $G_{ik}(\bm x, \bm x')$ by $G_{ik}(\bm x, \bm x')\equiv 8\pi u_i^{k}(\bm x, \bm x')$ is the corresponding Green's function. Therefore, as typical in the integral flow representations, any point $\bm x'$ at the surface induces the flow proportional to $u_i^{k}(\bm x, \bm x')$ at arbitrary point $\bm x$ in the fluid. The Green's function was first reported by Blake in Ref.~\cite{blake1971note}: 
\begin{eqnarray}&&\!\!\!\!\!\!\!\!\! 
G_{ik}(\bm x, \bm x')=\left(\frac{1}{q}-\frac{1}{R}\right)\delta_{ik}+\frac{q_iq_k}{q^3}-\frac{R_iR_k}{R^3}+2h(\delta_{k\beta}\delta_{\beta l}-\delta_{k3}\delta_{3l})\frac{\partial}{\partial R_l}\left(\frac{hR_i}{R^3}-\left(\frac{\delta_{i3}}{R}+\frac{R_iR_3}{R^3}\right)\right),\nonumber 
\label{Blake}
\end{eqnarray}
where $h\equiv z'$ is the height of the point-force singularity (source) above the plane,
$\bm q=(x-x', y-y', z-h)$, $R=(x-x', y-y', z+h)$ and $\beta$ stands for either $x$ or $y$ (with no summation over $\beta$). 
The far-field flow involves determining the asymptotic behavior of the Green's function at large distances from the source. Its leading-order asymptotic form for $|x'|\ll |x|$ and $|y'|\ll |y|$ is found by simply setting $x'=y'=0$. However, for $h\ll z$ we also need the higher-order term(s) in the Taylor series due to degeneracy $G_{ik}(h=0)=0$ (the flow is fully hindered by the no-slip plane) such that
\begin{eqnarray}&&\!\!\!\!\!\!\!\!\!
G_{ik}(\bm x, \bm x')\approx h\frac{\partial G_{ik}}{\partial h}(x'=y'=h=0), \label{aps}
\end{eqnarray}
where
\begin{eqnarray}&&\!\!\!\!\!
\frac{\partial G_{ik}}{\partial h}(x'\!=\!y'\!=\!h\!=\!0)\!=\!-2\frac{\partial}{\partial z}\left(\frac{\delta_{ik}}{r}\!+\!\frac{r_ir_k}{r^3}\right)
\!-\!2(\delta_{k\beta}\delta_{\beta l}\!-\!\delta_{k3}\delta_{3l})\frac{\partial}{\partial r_l}\left(\frac{\delta_{i3}}{r}\!+\!\frac{r_iz}{r^3}\right)\!=\!\nonumber\\&&\!\!\!\!\! 
-2\left(\frac{\delta_{i3}r_k\!+\!\delta_{k3}r_i}{r^3}\!-\!\frac{\delta_{ik}z}{r^3}\!-\!\frac{3zr_ir_k}{r^5}\right)
-2(\delta_{k\beta}\delta_{\beta l}-\delta_{k3}\delta_{3l})\left(\frac{\delta_{il}z+\delta_{3l}r_i}{r^3}-\frac{\delta_{i3}r_l}{r^3}-\frac{3r_lr_iz}{r^5}\right).\  \ \nonumber 
\end{eqnarray}
After some algebra the above result simplifies to a compact formula: 
\begin{eqnarray}&&\!\!\!\!\!\!\!\!\! 
\frac{\partial G_{ik}}{\partial h}(x'\!=\!y'\!=\!h\!=\!0)=\delta_{k\beta}\frac{12zr_{\beta}r_i}{r^5}.\label{simple}
\end{eqnarray}
Substituting Eqs.~(\ref{aps}) and (\ref{simple}) into Eqs.~(\ref{onl}) gives the flow away from the source for $|x'|\ll |x|$, $|y'|\ll |y|$ and $h\ll z$:  
\begin{eqnarray}&&\!\!\!\!\!\!\!\!\! 
v_i(\bm x)\approx -\frac{1}{8\pi}\frac{\partial G_{ik}}{\partial h}(x'\!=\!y'\!=\!h=\!0)
\left(\oint_{S} z\left(\delta_{kx}\delta_{zl}+\delta_{kz}\delta_{xl}\right)dS_l+M_k\right)=
-\frac{3z}{2\pi}\delta_{k\beta}\frac{r_{\beta}r_i}{r^5}\left(M_k+\delta_{kx} \oint_{S} zdS_z\right)
\,, 
\label{moment}
\end{eqnarray}
where we introduced the stress moment 
\begin{equation}
M_k\equiv \oint_{S} z \sigma_{kl}  dS_l\,. \label{Mk1}
\end{equation}
Using $\oint_{S} z dS_z=\int_V\nabla_i \cdot  (\delta_{iz} z)=V$, where $V$ is the particle volume, the Eq.~(\ref{moment}) reads 
\begin{eqnarray}&&\!\!\!\!\!\!\!\!\! 
\bm v(\bm x)\approx 
-\frac{3z\left(x(V+M_x)+yM_{y}\right)}{2\pi |\bm x|^5}\bm x\,. \label{farflow}
\end{eqnarray}
Adsorbed particle of an arbitrary shape induces universal far-field flow disturbance described by Eq.~(\ref{farflow}). The formula holds at distances from the particle that are much larger than its characteristic linear dimension. 

\subsection{Flow due to a sparse uniform distribution of particles} \label{unf}

The far flow given by Eq.~(\ref{farflow}) decays $\propto x^{-2}$ with the distance $x$ from the particle (compare to $x^{-1}$-decay for the Stokeslet in unbounded fluid). This is analogous to the spatial decay of the electric field induced by a point charge in electrostatics. It is well-known that when charges are uniformly distributed over an infinite plane, the total electric field that they induce is independent of the distance from the plane. Hence, one can anticipate that a uniform distribution of adsorbed particles will induce constant flow away from the plane. 

The Eq.~(\ref{farflow}) holds for the particle affixed to the plane at the  origin. However, due to invariance of the horizontal directions, the flow originated by the particle near the arbitrary point $(x', y')$ at the plane can be obtained by simply replacing $x$ with $x-x'$ and $y$ with $y-y'$ in Eq.~(\ref{farflow}). Thus, if identical particles are adsorbed uniformly over the plane with some number areal density $n$, the flow that they induce depends on the coordinate $z$ only and is given by (we use $\bm x'\equiv (x', y', 0)$ and $\bm r=(x, y, z)$):
\begin{eqnarray}&&\!\!\!\!\!\!\!\!\! v_i(z)=-\frac{3zn}{2\pi}\int \frac{(x-x')(V+M_x)+(y-y')M_{y}}{((x-x')^2+(y-y')^2+z^2)^{5/2}}(x_i-x'_i) dx'dy'
=-\frac{3zn}{2\pi }\int \frac{r_i\left(x(V+M_x)+yM_{y}\right)}{r^{5}} dx dy. \label{fda}
\end{eqnarray}
Here we assume that the particle areal number density $n$ is low enough, such that the net flow is given by the superposition of the individual contributions due to each particle. The ``low-surface coverage" assumption will be considered below. Eq.~(\ref{fda}) also assumes that the collection of particles can be considered as a smooth surface layer rather than the collection of discrete particles. This is valid provided that the characteristic inter-particle distance is much smaller than $z$ so that contributions of nearby particles are similar, i.e., $n^{1/2} z \gg 1$. 

The Eq.~(\ref{fda}) indicates that, as one may anticipate from symmetry, the vertical flow vanishes, $v_z(z)=0$. The nontrivial horizontal components of the flow are given by 
\begin{eqnarray}&&\!\!\!\!\!\!\!\!\! 
v_x(z)=-\frac{3zn(V+M_x)}{2\pi }\int \frac{x^2 dx dy}{(x^2+y^2+z^2)^{5/2}}=-n(V+M_x),\ \  v_y(z)=-nM_y. \label{fas}
\end{eqnarray}
Here and thereafter we use the identities 
\begin{eqnarray}&&\!\!\!\!\!
\int \frac{dxdy}{r^3}=\frac{2\pi}{z},\ \ \  \int \frac{dxdy}{r^5}=\frac{2\pi}{3z^3},\ \ \ 
\int \frac{x^2 dxdy}{r^5}=\int \frac{y^2 dxdy}{r^5}=\frac{1}{2}\int \frac{(r^2-z^2)dxdy}{r^5}=\frac{2\pi}{3z}. \label{ins}
\end{eqnarray}
The Eqs.~(\ref{fas}) are remarkably simple and they confirm that the net flow due to a layer of adsorbed particles does not decay away from the plain, in accord with the electrostatic analogy above. However, the details are different: in electrostatics the induced field is vertical, while the far-field flow is horizontal. The $x-$component of the flow indicates flow hindrance due to a layer of adsorbed particles. (It probably can be rigorously shown that the term $V+M_x$ is always positive, so that particles always hinder the flow). Although the flow disturbance does not decay with the distance from the plane, obviously the ratio of this constant flow and the unperturbed linear shear flow tends to zero away from the plane. On the other hand, the prospective transverse flow given by $v_y$, is a non-trivial leading-order contribution due to the adsorbed particles. Qualitatively, it is a result of hydrodynamic interaction of the shear flow with the asymmetric particles for which $M_y\neq 0$. 

To the leading order, there is no vertical component in the far-field flow. The $z-$component of the flow due to a single particle in Eq.~(\ref{farflow}) is odd in $x$ and in $y$. For instance, a spherical particle at the origin induces the vertical flow 
\begin{eqnarray}&&\!\!\!\!\!\!\!\!\! 
v_z(\bm x)\approx -\frac{3(V+M_x)}{2\pi}\frac{xz^2}{r^5}, \label{farflow0}
\end{eqnarray}
where $M_y=0$ by symmetry, see the proof in the next subsection. The flow is odd in $x$, being directed upward for $x<0$ and downward for $x>0$. For uniform surface distribution of spherical adsorbates, the contributions of individual particles compensate each other and the net vertical flow vanishes (the case of non-homogeneous distributions is considered in the next subsection).

The velocity field induced by a particle near the point $(x', y')$ at the plane, considered as a function of $x'$ and $y'$ varies over the length scale of $\mathcal{O}(z)$. This implies the condition of uniform distribution holds given that the number of particles within an arbitrary circle with radius $r\simeq z$ is about the same. For instance if the particles are distributed over a square lattice the result would hold at $z$ much larger than the lattice constant (and clearly much larger than the particle size).

\subsection{Far-field flow due to a spherical particle}

In this subsection we consider the far-field perturbation of the shear flow due to a spherical particle affixed to a plane. We use the spherical coordinate system $(r,\theta,\varphi)$ with origin at the sphere center. We shall demonstrate below that the only non-vanishing component of $M_k$ in (\ref{Mk1}) is $M_x$, which is computed explicitly using the above solution in tangent sphere coordinates. Using the analysis of Sec. \ref{s} one can show that
\begin{eqnarray}&&\!\!\!\!\!\!\!\!\! 
M_k=-\oint_{r=1} \! zP\cos\varphi\delta_{kr}dS+\oint_{r=1} \! z\tau_{kr}dS=-\pi \delta_{kx}\int_0^{\pi} \! zP\sin^2\theta d\theta+\oint_{r=1} \! z\tau_{kr}dS,\ \nonumber 
\end{eqnarray}
where $\tau_{il}=\partial_l v_i+\partial_i v_l$ are the components of the viscous stress tensor. Using the definitions of Sec.~\ref{s} we can write $\tau_{zz}$ and $\tau_{zy}$ in term of $U, V$ and $W$ in the cylindrical coordinates as 
\begin{eqnarray}&&\!\!\!\!\!\!\!\!\! \tau_{zz}=2\frac{\partial v_z}{\partial z}=2\frac{\partial W}{\partial z}\cos\varphi,\ \ \tau_{zx}=
\frac{\partial (W\cos\varphi)}{\partial x}+\frac{\partial v_x}{\partial z}=\cos^2\varphi\left(\partial_{\rho}W
+\partial_zU\right)+\sin^2\varphi\left(\frac{ W}{\rho}-\partial_zV\right),\nonumber\\&&\!\!\!\!\!\!\!\!\!
\tau_{zy}=\frac{\partial (W\cos\varphi)}{\partial y}+\frac{\partial ((U+V)\sin\varphi \cos\varphi)}{\partial z}=\sin\varphi \cos\varphi\left(\partial_{\rho}W-\frac{W}{\rho}+\frac{\partial (U+V)}{\partial z}\right). 
\end{eqnarray}
Then, integration over $\varphi$ yields 
\begin{eqnarray}&&\!\!\!\!\!\!\!\!\! 
M_z=\oint_{r=1} \! z\left(\cos\theta\tau_{zz}+\sin\theta\cos\varphi\tau_{zx}+\sin\theta\sin\varphi\tau_{zy}\right)dS=0.
\end{eqnarray}

Similarly, the stress components $\tau_{yy}$ and $\tau_{yx}$ read
\begin{eqnarray}&&\!\!\!\!\!\! \frac{\tau_{yy}}{2}\!=\!\frac{\partial v_y}{\partial y}\!=\!\sin\varphi \cos\varphi\frac{\partial(U\!+\!V)}{\partial y}\!+\!
(U\!+\!V)\frac{\cos\varphi(\cos^2\varphi\!-\!\sin^2\varphi)}{\rho}\!=\!\sin^2\varphi \cos\varphi\left(\frac{\partial(U\!+\!V)}{\partial \rho}\!-\!\frac{2(U\!+\!V)}{\rho}\right)\!+\!
\frac{\cos\varphi(U\!+\!V)}{\rho},\nonumber\\&&\!\!\!\!\!\!
\tau_{yx}=\sin\varphi\cos^2\varphi
\left(\partial_{\rho} U-\frac{2(U+V)}{\rho}\right)-\sin^3\varphi\partial_{\rho} V+
\sin\varphi\cos^2\varphi\partial_{\rho}(U+V)+(U+V)\frac{\sin\varphi(\sin^2\varphi-\cos^2\varphi)}{\rho},
\end{eqnarray}
and integration over $\varphi$ gives 
\begin{eqnarray}&&\!\!\!\!\!\!\!\!\! 
M_y=
\oint_{r=1} \! z\left(\cos\theta\tau_{yz}+\sin\theta\cos\varphi\tau_{yx}+\sin\theta\sin\varphi\tau_{yy}\right)dS=0.
\end{eqnarray}
We therefore conclude that for a sphere affixed to a plane the the only nontrivial component of $M_k$ is the $x$-component, i.e., $M_k=\delta_{kx} M$. 

Returning to the coordinate system with the origin at the point of contact, as used in previous subsection, it follows from Eq.~(\ref{farflow0}) that far from an adsorbed sphere the flow disturbance is given by
\begin{eqnarray}&&\!\!\!\!\!\!\!\!\! 
\bm v(\bm x)\approx -\alpha\frac{xz\bm r}{r^5},\ \ \alpha\equiv 2+\frac{3M}{2\pi}. \label{aslp}
\end{eqnarray}
The formula for pressure in the far field limit is obtained by solving the Stokes equation with the above flow,
\begin{eqnarray}&&\!\!\!\!\!\!\!\!\! 
\partial_i p=\nabla^2 v_i\approx 
-2\alpha\left(\frac{z\delta_{ix}+x\delta_{iz}}{r^5}-\frac{5 xzr_i}{r^7}\right)\quad \Rightarrow \quad p\approx -2\alpha\frac{xz}{r^5}.\label{farp}
\end{eqnarray}
Note that at large distances there is a simple relation between pressure and vertical velocity, $v_z\approx zp/2$. 

To determine $M$ we compare the above asymptotic formula for the far-firld pressure with the full solution given by the first of Eqs.~(\ref{oneil}). Eq.~(\ref{farp}) gives
\begin{eqnarray}&&\!\!\!\!\!\!\!\!\! 
rP\approx -2\alpha\frac{\sin\theta\cos\theta}{r^2}=-\frac{\alpha\eta\xi}{2}.
\end{eqnarray}
Using $\sqrt{\xi^2\!+\!\eta^2}=2/r$ and Eqs.~(\ref{oneil}) we find that at $r\to \infty$ or $\sqrt{\xi^2\!+\!\eta^2}\to 0$
\begin{eqnarray}&&\!\!\!\!\!\!\!\!\! 
-\frac{\alpha\eta\xi}{2}\! \sim\!2\int_0^{\infty} \!\!\left(B(k)\cosh(k\xi)\!+\!C(k)\sinh(k\xi)\right)J_{1}(k\eta) dk\sim \eta\int_0^{\infty} B(k) k dk+\eta\xi\int_0^{\infty} C(k) k^2 dk+\ldots,\label{doa}
\end{eqnarray}
where we used the series formula for $J_1$ and dots stand for higher-order terms. Substituting the expression for $B(k)$ from Eqs.~(\ref{allcoef}) it can be shown that the first term in the RHS of Eq.~(\ref{doa}) is zero:
\begin{eqnarray}&&\!\!\!\!\!\!\!\!\! 
\int_0^{\infty} B(k) k dk=\int_0^{\infty} \left(k^2A''+kA'-A\right) dk=0.
\end{eqnarray}
Using the second term in the RHS of Eq.~(\ref{doa}) it follows that:
\begin{eqnarray}&&\!\!\!\!\!\!\!\!\! 
\frac{\alpha}{2}\!=\!-\int_0^{\infty}\!\! C(k) k^2 dk\!=\!\int_0^{\infty} (kA''\!-\!A') (k^2\coth k\!-\!k) dk\!=\!
\int_0^{\infty} A\left(8k\coth k-\frac{7k^2}{\sinh^2k}-3+\frac{2k^3\cosh k}{\sinh^3k}\right)dk. \label{z-moment1}
\end{eqnarray}
Using the above numerical solution for $A(k)$ (see Fig.~\ref{fig:Ak}) yields $\alpha=21.1745$. Finally, it follows from Eq.~(\ref{aslp}) that 
\begin{equation}
M=\oint_{r=1} z \sigma_{xl}  dS_l = \frac{4 \pi}{3}\left(\frac{\alpha}{2}-1\right)\simeq 40.159\,, \label{Mx}    
\end{equation}
This result can also be obtained using the direct integration of the tractive force over the particle surface at $r=1$ (see Appendix~\ref{app:A}).  

We now shall go back the underlying assumptions of the analysis in this subsection. The Eq.~(\ref{fas}) indicates that a dilute uniform layer of spherical particles adsorbed at the plane surface induces a backflow $-n (4\pi/3+M)$ with $M\simeq 40.159$ and no lateral flow, as $M_y=0$. It can be readily seen, that the contribution proportional to the particles volume, $\propto n V$ is small in comparison to that related to the dipole stress moment $\propto nM$. The disturbance to the linear shear flow created due to the layer of spherical particles obeys Eqs.~(\ref{eqv}) with the no-slip boundary conditions holding on each sphere. The flow given by superposition of individual contributions induced by each adsorbate is a good approximation provided that the flow disturbance in the near vicinity of a single particle due to rest of the particles (i.e., the hydrodynamic interaction) is negligible. Since the far-field flow due to each sphere holds at distances much larger than the particle radius, irrespective of $z$, then $-n(4\pi/3+M)\simeq -nM$ provides an accurate approximation for the flow induced by the rest of the particles. Therefore, the low-surface coverage assumption holds if $n M\ll 1$. 
In the dimensional form, this condition reads $n a^2\ll M^{-1}$.
Notice that the extra condition on the vertical distance due to the approximation of smooth layer in subsection \ref{unf} implies $n^{1/2} z\ll 1$.

Finally, it is important to stress that for asymmetric particles, however, $M_y$ may not vanish and they may generate a constant flow that does not decay away from the plane, in the direction transverse to the direction of the shear flow.

\begin{acknowledgments}
This work was supported, in part, by the Israel Science Foundation (ISF) via the grant No. 1744/17. We thank Boris Rubinstein for technical assistance with Fig.~\ref{fig:flow}.  A.M.L. also acknowledges the support of the David T. Siegel Chair in Fluid Mechanics. 
\end{acknowledgments}

\newpage
\begin{appendices}
\appendix 

\section{Direct derivation of the force, torque and stress moment due to O'Neill~\cite{o1968sphere}} \label{app:A}
\setcounter{equation}{0}
\renewcommand{\theequation}{A\arabic{equation}}

The original derivations of the force and torque the shear flow exerts on the particle rigidly attached to a plane in Ref.~\cite{o1968sphere} relies on integration of traction over the particle surface at $r=1$. 
The relevant components of the traction in the cylindrical coordinates 
$\{\rho,\varphi, z\}$ reads:
\begin{eqnarray}
f_i  \equiv \sigma_{ik} n_k &=& -p \bm n + (\bm e_\rho \cdot \bm n) \hat{\bm e}_\rho \tau_{\rho\rho} + (\hat{\bm e}_\rho \cdot \bm n) \hat{\bm e}_\varphi \tau_{\rho\varphi}+(\hat{\bm e}_\rho \cdot \bm n) \hat{\bm e}_z \tau_{\rho z}+\nonumber \\
&& (\hat{\bm e}_z \cdot \bm n) \hat{\bm e}_\rho \tau_{z\rho}+(\hat{\bm e}_z \cdot \bm n) \hat{\bm e}_\varphi \tau_{z\varphi}+(\hat{\bm e}_z \cdot \bm n) \hat{\bm e}_z \tau_{zz}, \label{stress}
\end{eqnarray}
Thus, the traction $f_x$ reads:
\[
f_x=\sigma_{xk} n_k=-p (\bm n \cdot \hat{\bm x})+ (\hat{\bm e}_\rho \cdot \bm n) (\tau_{\rho\rho} {\hat{\bm e}}_\rho +\tau_{\rho\varphi} \hat{\bm e}_\varphi) \cdot \hat{\bm x}+(\hat{\bm e}_z \cdot \bm n) (\tau_{z\rho} {\hat{\bm e}}_\rho +\tau_{z\varphi} \hat{\bm e}_\varphi) \cdot \hat{\bm x}\,,
\]
Introducing the spherical coordinates $(r, \theta, \varphi)$ with the origin at the sphere center, such that $\bm n=\hat{\bm e}_r$, and using $\hat{\bm x} \cdot \bm n=\sin{\theta}\cos{\varphi}$, $\hat{\bm e}_\rho \cdot \bm n=\sin{\theta}$, $\hat{\bm e}_z\cdot \bm n=\cos{\theta}$, $\hat{\bm e}_\rho\cdot \hat{\bm x}=\cos{\varphi}$ and $\hat{\bm e}_\varphi\cdot \hat{\bm x}=-\sin{\varphi}$ we readily obtain:
\begin{equation}
f_x=- p \sin\theta \cos\varphi+\sin\theta \cos\varphi\, \tau_{\rho\rho}-\sin\theta\sin\varphi\, \tau_{\rho\varphi}+\cos\theta\cos\varphi\, \tau_{z\rho}-\cos\theta\sin\varphi\, \tau_{z\varphi}, \label{fx}    
\end{equation}
We next substitute the viscous stress components 
\begin{eqnarray}
    && \tau_{\rho\rho}=2\frac{\partial v_\rho}{\partial \rho}, \quad
    \tau_{\rho\varphi}= \rho\frac{\partial}{\partial \rho}\left(\frac{v_\varphi}{\rho}\right)+\frac{1}{\rho}\frac{\partial v_\rho}{\partial \varphi}, \nonumber \\
    && \tau_{z\rho}= \frac{\partial v_z}{\partial \rho}+\frac{\partial v_\rho}{\partial z}, \quad 
    \tau_{z\varphi}= \frac{\partial v_\varphi}{\partial z}+\frac{1}{\rho}\frac{\partial v_z}{\partial \varphi}, \nonumber
\end{eqnarray}
making use of the ansatz in Eq.~(\ref{ca})
where $P=2Q,$ $U$, $V$ and $W$ are functions of $\rho$ and $z$ only.
Using  the boundary conditions at the particle surface, $U=-V$, which together with continuity equation (\ref{incomp}) gives
after some algebra the traction:
\begin{eqnarray}
f_x &=&\sin{\theta} \left[ -2Q \cos^2{\varphi}+2\frac{\partial U}{\partial \rho} \cos^2{\varphi} -\frac{\partial V}{\partial \rho} \sin^2{\varphi} \right]+ \nonumber \\
&& \qquad \cos\theta\left[\frac{\partial U}{\partial z}\cos^2{\varphi} - \frac{\partial V}{\partial z} \sin^2{\varphi}+ \frac{\partial W}{\partial \rho} \cos^2{\varphi}\right]\,. \label{fx1}    
\end{eqnarray}
Integrating $f_x$ in (\ref{fx1}) over the angle $\varphi$, as the integrals of $\cos^2\varphi$ and $\sin^2\varphi$ both give $\pi$, we obtain the force in the form
\[
F_x=\oint_{r=1} f_x dS=\pi \int\limits_0^\pi \left(\left[ -2 Q +2\frac{\partial U}{\partial \rho} -\frac{\partial V}{\partial \rho}\right] \sin{\theta}+ \left[\frac{\partial U}{\partial z} - \frac{\partial V}{\partial z} + \frac{\partial W}{\partial \rho}\right] \cos{\theta} \right) \sin\theta d\theta\,.
\]
Replacing the integration variable with $\beta=\pi-\theta$, we readily obtain the unnumbered equation on top of the right column on p.~1296 in \cite{o1968sphere}:
\begin{equation}
    F_x=-\pi \int\limits_0^\pi \left(\left[ 2 Q -2\frac{\partial U}{\partial \rho} +\frac{\partial V}{\partial \rho}\right] \sin{\beta}+ \left[\frac{\partial U}{\partial z} - \frac{\partial V}{\partial z} + \frac{\partial W}{\partial \rho}\right] \cos{\beta} \right) \sin\beta d\beta\,. \label{Fx1o}
\end{equation}

Substituting $U=\rho Q+(\chi+\psi)/2$,  $V=(\chi-\psi)/2$ and $W=zQ+\phi$ together with 
\[
\sin{\beta}=\frac{2\eta}{1+\eta^2}\,,\quad \cos\beta=1-\frac{2}{1+\eta^2}\,, \quad d\beta=-\frac{2d\eta}{1+\eta^2},
\]
into (\ref{Fx1o}) yields after some algebra the Eq.~4.1 of O'Neill \cite{o1968sphere}:
\begin{equation}
    F_x=-2 \pi \int\limits_0^\infty \left\{ \rho\frac{\partial Q}{\partial \xi}-Q \frac{\partial \rho}{\partial \xi}+ \frac{\partial \psi}{\partial \xi} \right\}_{\xi=1}\, \frac{\eta d\eta}{1+\eta^2}\,. \label{Fx2o}
\end{equation}
Substituting the solution for $Q$ from (\ref{oneil}) into the fist term in the figure brackets in (\ref{Fx2o}), integrating over $\eta$ by using the definite integrals involving Bessel functions (see Eqs.4.3--4.4 in \cite{o1968sphere}), yields:
\[
{\mathcal T}_1=\int\limits_0^\infty \left(\rho\frac{\partial Q}{\partial \xi}\right)_{\xi=1}\, \frac{\eta d\eta}{1+\eta^2}=B k e^{-k} \left(\frac{\cosh{k}}{3}+\sinh{k}\right) + C k e^{-k} \left(\frac{\sinh{k}}{3}+\cosh{k}\right)\,,  
\]
Analogously, the second term in the figure brackets in (\ref{Fx2o}) yields
\[
{\mathcal T}_2=\int\limits_0^\infty \left(-Q\frac{\partial \rho}{\partial \xi}\right)_{\xi=1}\, \frac{\eta d\eta}{1+\eta^2}=\frac{2}{3} k e^{-k} (B \cosh{k}+ C\sinh{k})\,.
\]
Lastly, substituting the corresponding solutions for $\psi$ in (\ref{oneil}), the 3rd term in the figure brackets in (\ref{Fx2o}) produces
\[
{\mathcal T}_3=\int\limits_0^\infty \left(\frac{\partial \psi}{\partial \xi}\right)_{\xi=1}\, \frac{\eta d\eta}{1+\eta^2}=
(D \sinh{k}+E \cosh{k}) e^{-k}+ (D \cosh{k}+E \sinh{k}) e^{-k}=D+E\,.
\]
Summing up all three contributions we readily find ${\mathcal T}_1+{\mathcal T}_2+{\mathcal T}_3=k (B+C)+D+E$ and 
therefore 
\begin{equation}
F_x=-2\pi \int\limits_0^\infty \left(k [B+C]+D+E\right)\, dk\,. \label{Fx3o}    
\end{equation}

Using Eq.~(\ref{allcoef}) and expressing all coefficients in terms of $A(k)$ the integrand in (\ref{Fx3o}) simplifies to  
\[
k (B+C)+D+E=A''k^2(K+1)-4k(\coth{k}-1)+2kA'=k^2 A'' K-4k(\coth{k}-1)+(k^2 A')'\,, 
\]
where as before $K=k^{-1}-\coth{k}$. Since $(k^2 A')'$ has no contribution to the integral, the resulting expression for the force reads
\begin{equation}
F_x=2\pi \int\limits_0^\infty \left\{4k(\coth{k}-1)-k^2 A'' K \right\} dk\, \label{Fx4o}
    \end{equation}
which is identical to Eq.~(\ref{Fx}), which was derived using an alternative (and simpler) approach.   


The viscous torque the flow exerts on the particle (about its center) reads:
\begin{equation}
    L_i=\epsilon_{ijl} \oint_{r=1} x_j \sigma_{lk} dS_k
\end{equation}
and rewritten in spherical coordinates 
\begin{equation}
   L_y= \oint_{r=1} \left( \cos{\theta}\, \sigma_{xk} -\sin{\theta} \cos{\varphi}\, \sigma_{zk} \right) dS_k\,, \label{torque}
\end{equation}
where $\theta$ and $\varphi$ are the spherical angles. Following the derivation similar to that of the force, the expression for torque in Eq.~(\ref{torque}) reduces after some algebra to 
\begin{equation}
L_y=-F_x+8\pi\int_0^\infty k A(k) dk\,, \label{Lyo}
\end{equation}
which is identical to Eq.~(\ref{Ly}).


Finally, we compute the stress moment $M_x$ controlling the far-field flow disturbance due to an adsorbed spherical particle using the direct integration of over the particle surface:
\begin{equation}
M\equiv M_x = \oint_{r=1} z \sigma_{xk} n_k dS\,. \label{Mxo}
\end{equation}
The traction $f_x=\sigma_{xk} n_k$ is given by Eq.~(\ref{fx1}) and using an approach similar to that applied to derive Eq.~(\ref{Fx2o}), we readily obtain:
\begin{equation}
M =  -2 \pi \int\limits_0^\infty \left\{ \rho\frac{\partial Q}{\partial \xi}-Q \frac{\partial \rho}{\partial \xi}+ \frac{\partial \psi}{\partial \xi} \right\}_{\xi=1}\, \frac{2\eta\, d\eta}{(1+\eta^2)^2}\,.
\,, \label{Fx5o}
\end{equation}
where we substituted $z=2/(1+\eta^2)$ at the particle surface, $\xi=1$.

Proceeding as before, we  substitute the solution for $Q$ from (\ref{oneil}) into the first term within the figure brackets under the integral in Eq.~(\ref{Fx5o}), integrate first over $\eta$ to obtain:
\[
\mathcal{I}_1=\int\limits_0^\infty \left(\rho\frac{\partial Q}{\partial \xi}\right)_{\xi=1}\, \frac{2\eta\, d\eta}{(1+\eta^2)^2}= \frac{2}{15} k (1+k) e^{-k}  (B\cosh{k}+C \sinh{k}) + \frac{2}{3} k^2 e^{-k} (B \sinh{k}+ C \cosh{k})\,,  
\]
where we used the following integrals (see also Eqs.4.3--4.4 in \cite{o1968sphere}):
\[
\int_0^\infty \frac{\eta^2 J_1(k\eta)} {(1+\eta^2)^{7/2}} d\eta=\frac{1}{15} k (1+k) e^{-k}\,,\qquad \int_0^\infty \frac{\eta^2 J_1(k\eta)} {(1+\eta^2)^{5/2}} d\eta=\frac{1}{3} k e^{-k}\,.
\]
Analogously, the second term under the integral sign in (\ref{Fx5o}) produces
\[
\mathcal{I}_2=\int\limits_0^\infty \left(-Q\frac{\partial \rho}{\partial \xi}\right)_{\xi=1}\, \frac{2\eta d\eta}{(1+\eta^2)^2}=\frac{4}{15} k (1+k) e^{-k} (B \cosh{k}+ C \sinh{k})\,.
\]
Lastly, substituting the solution for $\psi$ from Eq.~(\ref{oneil}) into the 3rd term under the integral sign in (\ref{Fx5o}) results in
\[
\mathcal{I}_3=\int\limits_0^\infty \left(\frac{\partial \psi}{\partial \xi}\right)_{\xi=1}\, \frac{2 \eta d\eta}{(1+\eta^2)^2}= \frac{2}{3} (1+k) e^{-k}(D\cosh{k}+E \sinh{k}) + 2 k e^{-k} (D \sinh{k}+E \cosh{k})\,.
\]
Summing up $\mathcal{I}_1$, $\mathcal{I}_2$ and $\mathcal{I}_3$, substituting the functions $B, C, D$ and $E$ via $A$ from Eq.~(\ref{allcoef}) and integrating the resulting (lengthy) expression containing $A, A'$ and $A''$ over $k$ from 0 to infinity finally gives
\begin{equation}
M=\oint_{r=1} z \sigma_{xk} dS_k
\simeq 40.1594\,, \label{Mxo}
\end{equation}
This value is an excellent agreement with the result $M\simeq 40.159$ in Eq.~(\ref{Mx}) obtained using the far-field flow analysis .


\end{appendices}


\end{document}